\documentclass[journal]{IEEEtran}

\IEEEoverridecommandlockouts
\makeatletter
\def\ps@headings{%
\def\@oddhead{\mbox{}\scriptsize\rightmark \hfil \thepage}%
\def\@evenhead{\scriptsize\thepage \hfil \leftmark\mbox{}}%
\def\@oddfoot{}%
\def\@evenfoot{}}
\makeatother
\usepackage{subfigure}
\usepackage{colortbl}
\usepackage{bm}

\usepackage{algpseudocode}  
\renewcommand{\algorithmicrequire}{\textbf{Initialization:}}  

\usepackage{graphicx}  
\usepackage{url}       

\usepackage{amsmath}   
\usepackage{cite}

\usepackage{amsfonts,amssymb}
\usepackage{cite}
\usepackage{stfloats}

\usepackage{cases}
\usepackage{algorithm}
\usepackage{multirow}
\usepackage{algorithmicx}
\usepackage{stfloats}
\usepackage{epstopdf}

\usepackage{xcolor}
\usepackage{xpatch}
\makeatletter
\def\changeBibColor#1{%
	\in@{#1}{
	}
	\ifin@\color{blue}\else\normalcolor\fi
}
\xpatchcmd\@bibitem
{\item}
{\changeBibColor{#1}\item}
{}{\fail}
\xpatchcmd\@lbibitem
{\item}
{\changeBibColor{#2}\item}
{}{\fail}

\newcommand{\Rmnum}[1]{\expandafter\@slowromancap\romannumeral #1@}

\makeatother

\UseRawInputEncoding

\newtheorem{theorem}{{Theorem}}

\newcommand{\ls}[1]
    {\dimen0=\fontdimen6\the\font
     \lineskip=#1\dimen0
     \advance\lineskip.5\fontdimen5\the\font
     \advance\lineskip-\dimen0
     \lineskiplimit=.9\lineskip
     \baselineskip=\lineskip
     \advance\baselineskip\dimen0
     \normallineskip\lineskip
     \normallineskiplimit\lineskiplimit
     \normalbaselineskip\baselineskip
     \ignorespaces
    }

\begin{document}

\title{Self-Sustainable Active Reconfigurable Intelligent Surfaces for Anti-Jamming in Wireless Communications}
\vspace{10pt}

%
%
%

\author{\IEEEauthorblockN{Yang Cao, \emph{Graduate Student Member}, \emph{IEEE}, Wenchi Cheng, \emph{Senior Member}, \emph{IEEE}, Jingqing Wang, \emph{Member}, \emph{IEEE}, and Wei Zhang, \emph{Fellow}, \emph{IEEE}}\\[0.2cm]
	\vspace{-5pt}
	
	
	\vspace{-25pt}
	
	\thanks{Part of this work was presented in IEEE International Conference on Communications, 2022\cite{Cao1}.
		
		Yang Cao, Wenchi Cheng and Jingqing Wang are with the State Key Laboratory of Integrated Services Networks, Xidian University, Xi'an,
		710071, China (e-mails: caoyang@stu.xidian.edu.cn, wccheng@xidian.edu.cn, wangjingqing00@gmail.com).
		
		Wei Zhang is with the School of Electrical Engineering and Telecommunications, University of New South Wales, Sydney,
		NSW 2052, Australia (e-mail: w.zhang@unsw.edu.au).}
}

\maketitle

\begin{abstract}

  Wireless devices can be easily attacked by jammers during transmission, which is a potential security threat for wireless communications. Active reconfigurable intelligent surface (RIS) attracts  considerable attention and is expected to be employed in anti-jamming systems for secure transmission to significantly enhance the anti-jamming performance. However, active RIS introduces external power load, which increases the complexity of hardware and restricts the flexible deployment of active RIS. To overcome these drawbacks, we design a innovative self-sustainable structure in this paper, where the active RIS is energized by harvesting energy from base station (BS) signals through the time dividing based simultaneous wireless information and power transfer (TD-SWIPT) scheme. Based on the above structure, we develop the BS harvesting scheme based on joint transmit and reflecting beamforming  with the aim of maximizing the achievable rate of active RIS-assisted system, where the alternating optimization (AO) algorithm based on stochastic successive convex approximation (SSCA) tackles the nonconvex optimization problem in the scheme. Simulation results verified the effectiveness of our developed BS harvesting scheme, which can attain higher anti-jamming performance than other schemes when given the same maximum transmit power.

\end{abstract}


\begin{IEEEkeywords}
	
Active reconfigurable intelligent surface, self-sustainable, anti-jamming, alternating optimizatio, stochastic successive convex approximation.
\end{IEEEkeywords}

\section{Introduction}
 \IEEEPARstart{A}{s} the emerging of diverse wireless communications services over the fifth-generation (5G) wireless networks, there has been a surge in wireless equipments connecting to the network \cite{Emergency, mmWave1}. The open and broadcast properties of wireless channels render the communications among legitimate wireless devices vulnerable to security threats, such as active jamming and passive eavesdropping. Jamming, which involves the interruption or blocking of communication among wireless devices by attackers with jamming signals, can be performed at different levels of the communication system\cite{Anti-Jamming}, which exacerbates the communication security risks of wireless devices. With the massive number of connected wireless devices, jamming becomes more easier and it is imperative to cope with malicious jamming against wireless communication to meet the requirements of security.

 To counter jamming attacks, several anti-jamming technologies have been investigated, such as mode frequency hopping (mode-FH), dynamic spectrum anti-jamming (DSAJ), and so on \cite{DSSS, Dynamic-Spectrum}. However, numerous relays and antennas are inevitably utilized in current technologies, resulting in excessive hardware costs and energy consumption. In addition, traditional anti-jamming technologies fail to respond to the diversified and intelligent development of malicious jammers. In order to overcome the above challenges, it is imperative to develop effective anti-jamming solutions with low power consumption. 
 
 Recently, reconfigurable intelligent surface (RIS) has set off an upsurge in the field of secure wireless communication. RIS consists of a number of reflecting elements (REs) and the controller. REs allow the phase or amplitude of the electromagnetic signals to be intentionally adjusted. Thus, RIS can boost or weaken the incident signals, thereby improving the performance of wireless communications \cite{Precoding-Designs}. Due to this distinctive characteristic, RIS paves a new way for secure wireless communication. The authors of \cite{Enabling} studied the single-user and single-eavesdropper RIS-assisted communication system, and showed that system security can benefit from RIS. The secure multiple-input and multiple-output (MIMO) system with the assistance of RIS was further studied to increase the secrecy rate \cite{MIMO_SEC}. Apart from the anti-eavesdropper wireless communications, RIS also can used to anti-jamming wireless communications. The authors of \cite{Jamming-Mitigation} devised an successive convex approximation (SCA) based anti-jamming scheme to strengthen legitimate transmissions and resist jamming signals in the single RIS-assisted air-to-ground network. The author of \cite{fast} adopted a fast reinforcement learning scheme based on jointly optimizing the reflecting beamforming of RIS and the transmit beamforming of base station (BS) to boost the anti-jamming capability of single RIS-aided system. 

However, there are some urgent challenges in RIS-assisted anti-jamming secure wireless communications yet to be addressed. For example, the receive signals suffer from double-fading attenuation in RIS-assisted secure wireless communications to result in small SINR at the receiver, since RIS is considered to be passive and only reflects the incident signals without extra energy consumption \cite{Versus-Passive}. Double-fading means that the cascade channel constructed by RIS subjects the passing signal suffer massive attenuation twice, leading to smaller performance gain in comparison to the attenuation without the RIS. Nowadays, a novel paradigm termed ``active RIS" is proposed to restrain double-fading attenuation with a unique structure, where each RE is equipped with an additional power load. Therefore, the phase and amplitude of the weak incoming signal are allowed to be reshaped by the active RIS at low power consumption. In contrast to relays or repeaters, active RIS, while consuming additional power, maintains the advantage of obtaining sufficient signal amplification without a high-consumption radio frequency link. The authors of \cite{Multiuser-MISO} verified that active RIS has a performance advantage over conventional passive RIS in MISO systems. The authors of \cite{Aided-Wireless} investigated the secure communication system with the aid of active RIS, showing that a few REs allows for significant level of system performance. An efficient scheme was proposed to address the non-convex secrecy rate optimization in active RIS-assisted secure communication scenario \cite{Aided-Secure-Tran}. However, the external power load at the active RIS contributes to the hardware complexity and restricts the flexible deployment of active RIS.

 To address the above concerns, we design a innovative self-sustainable structure for the active RIS, which is energized by harvesting energy from BS signals through the time dividing based simultaneous wireless information and power transfer (TD-SWIPT) scheme. Based on the aforementioned structure, we develop a joint transmit and reflecting beamforming based BS harvesting scheme to maximize achievable rate of the self-sustainable active RIS-assisted multiuser anti-jamming system. We then adopt the stochastic successive convex approximation (SSCA) based alternating optimization (AO) algorithm to address the nonconvex problem of maximizing the achievable rate.. Besides, taking into account the mobile feature of the UE, we derive random waypoint (RWP) based Nakagami-$m$ fading channel model and apply it for numerical simulations. Finally, numerical results demonstrate that our developed BS harvesting scheme achieves significant and effective anti-jamming performance.  When fewer REs are deployed, the jamming suppression capability of the BS harvesting scheme still outperforms other schemes under the same transmit power consumption. In addition, we analyze the influence of other factors for the anti-jamming system to validate the benefits of utilizing active RIS to strengthen wireless communication performance in jamming assault scenarios.


\begin{figure*}[t!]
	\vspace{-10pt}
	\centering
	\includegraphics[scale=0.13]{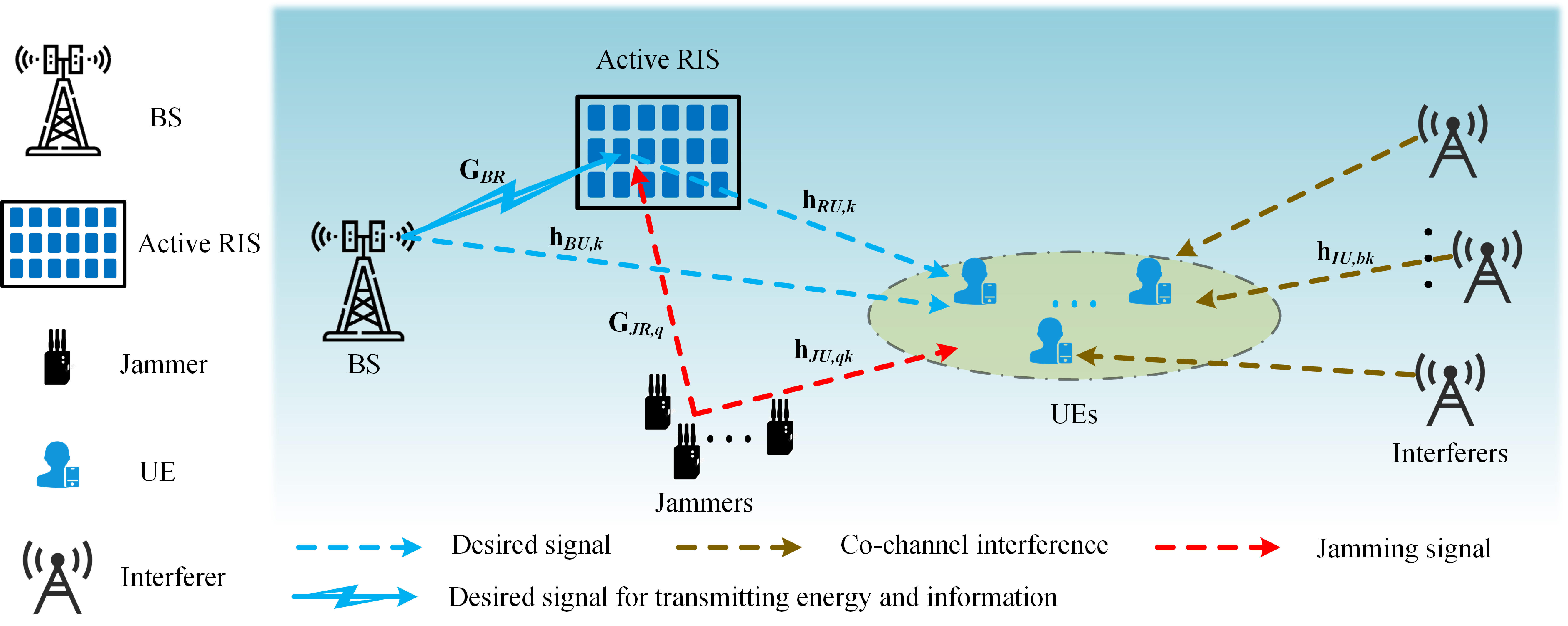}
	\vspace{-10pt}
	\caption{The active RIS-assisted anti-jamming communication system.}
	\vspace{-15pt}
	\label{fig:shiyitu}
\end{figure*}

  This paper is scheduled as follows. Section~\ref{sec:System_model} summarizes the anti-jamming system model including channel model and signal model. Section~\ref{sec:Problem_formulation} investigates the principle of the active RIS, and formulates the non-convex  problem to maximize the achievable rate. In Section~\ref{sec:AO}, an efficient SSCA-based AO algorithm is developed to tackle the non-convex problem for maximizing the achievable rate. In Section~\ref{sec:simulation}, numerical simulation are used to analyze the jamming suppression capability. This paper concludes with Section~\ref{sec:Conclusion}.

  \textit{Notations}: Boldface straight lower-case and upper-case letters denote vectors and matrices, respectively. The conjugate transpose, norm, transpose, diagonal matrix, and trace operation are denoted by $(\cdot)^H$, $\lVert\cdot\rVert$, $(\cdot)^T$, $\mbox{diag}(\cdot)$, and $\mbox{tr}(\cdot)$, respectively. The notation $\mathbb{C}^{x\times y}$ and $\lvert\cdot\rvert$ denote the space of $x\times y$ complex matrix and the absolute value of a scalar. $\boldsymbol{\rm T} \succeq \boldsymbol{\rm 0}$ represents that matrix $\boldsymbol{\rm T}$ denotes the positive semi-definite matrix. $\boldsymbol{\rm I}$ is an identity matrix with proper dimension.



\section{System Model}
\label{sec:System_model}

  In this paper, we consider the self-sustainable active RIS-assisted anti-jamming system, where the BS communicates with UEs aided by the active RIS. Fig.~\ref{fig:shiyitu} draws the anti-jamming system, consisting of one BS with $N$ antennas, $Q$ malicious jammers with $N_{Jam}$ antennas, one active RIS with $M$ REs, $B$ interferers with $N$ antennas, and $K$ single antenna UEs. Each RE of the active RIS is supported by the rechargeable battery in the anti-jamming system, thus realizing the joint dynamic configuration of phase and amplitude of REs. Therefore, the double fading of the cascaded channels can be significantly mended by the active RIS. Noteworthy is that the energy consumption required for the self-sustained active RIS is supplied by the BS transmitted signals using TD-SWIPT scheme. This novel RIS structure ensures the energy self-sufficiency of RIS and the communication quality of UEs.

  The BS transmits legitimate messages to UEs with the assistance of the self-sustainable active RIS. Malicious jammers are considered to be located near the UE and send jamming signals in an attempt to block the legitimate receptions of UEs. Specifically, the operation of jammers is divided into two stages: perception and jamming. Firstly, the intelligent malicious jammers observe the legitimate communication of UEs with general software radio peripherals. Jammers keep quiet when the channels related to UEs are idle. Then, they send out jamming signals with a delay of less than 30 $\mu s$ \cite{under-Broadband} immediately after detecting the communication signals. Malicious jammers launch jamming attacks to interfere with the UE, thus significantly resulting in legitimate communication failure. In addition, it is inevitable that interferers, transmit the same frequency signals to the UE, leading to co-channel interference. The active RIS is employed to restrain jamming and co-channel interference by modifying incident signals. Thus, the deployment of the active RIS provides additional links to resist jamming from jammers and interference from interferers, thus improving the communication performance of UEs.
  
 \vspace{-4mm}

  \subsection{Channel Model}
  \label{subsec:Channel_model} 
  
  Let the RE of the active RIS set, the jammer set, the UE set, and the interferer set be denoted by  $\mathcal{M}=\{1,2,\cdots,M\}$, $\mathcal{Q}=\{1,2,\cdots,Q\}$, $\mathcal{K}=\{1,2,\cdots,K\}$, and $\mathcal{B}=\{1,2,\cdots,B\}$, respectively. Let  $\boldsymbol{\rm G}_{BR} \in \mathbb{C}^{M\times N}$, $\boldsymbol{\rm h}_{BU,k}^H \in \mathbb{C}^{1\times N}$, $\boldsymbol {\rm h}_{RU,k}^H \in \mathbb{C}^{1\times M}$, $\boldsymbol {\rm h}_{JU,qk}^H \in \mathbb{C}^{1\times N_{Jam}}$, $\boldsymbol {\rm G}_{JR,q}^H \in \mathbb{C}^{M\times N_{Jam}}$, and $\boldsymbol {\rm h}_{IU,bk}^H \in \mathbb{C}^{1\times N}$ represent the channel coefficients between the BS and the active RIS, between the BS and the $k$th UE, between the active RIS and the $k$th UE, between the $q$th jammer and the $k$th UE, between the $q$th jammer and the active RIS, and between the $b$th interferer and the $k$th UE, respectively, where $b \in \mathcal{B}$, $k \in \mathcal{K}$, and $q \in \mathcal{Q}$. The phase and amplitude of the weak incoming signal are allowed to be reshaped by the active RIS, which is attributable to the external energy supply. The reflection coefficient matrix at active RIS is denoted as $\boldsymbol {\rm \Theta}= {\rm diag}(\boldsymbol{\rm {\theta}})\in \mathbb{C}^{M\times M}$, where $\boldsymbol{\rm {\theta}} = [\theta_{1},\theta_{2},\cdots,\theta_{M}]^T$. The notation $\theta_{m}$ represents the reflection coefficient between the $m$th RE of the active RIS, which is written as follows:
  \begin{equation}
  {\theta}_{m} = A_{m} e^{j\phi_{m}}, ~\forall m\in \mathcal{M},
  \end{equation} 
  where $A_{m} \in [0,A_{\max}], (A_{\max}\ge 1)$ indicates the amplitude with $A_{\max}$ being the maximum amplified amplitude. The notation $\phi_{m} \in [0,2\pi)$ represents the phase.

  
  
  In this paper, based on the existing channel estimation technology \cite{PARAFAC}, cooperating BS and UEs can easily track each other channel state information (CSI). Thus, the CSI of channels among the BS, the active RIS, and UEs is considered completely available. However, malicious jammers and interferers never cooperate with BS and UEs for CSI acquisition. Therefore, we assume that the imperfect CSI of jammers and interferers are available. In addition, we assume that wireless channels among the BS, the active RIS, UEs, jammers and interferers adopt the RWP-based Nakagami-$m$ fading channel model. All of the aforementioned channels is assumed to be modeled with quasi-static flat fading.

  The RWP model is a straightforward stochastic model for evaluating the mobility of various wireless systems, which describes the random movement of the UE within a given area \cite{Outage-BER}. The RWP model in this paper is used to model the behavior pattern of UEs moving randomly in the BS coverage area. Moreover, taking into account the generalization of the channel, we set the channel associated with the UEs to follow the Nakagami-$m$ fading model. Combining the RWP model and Nakagami-$m$ fading model, we have the following theorem.

%
  \begin{theorem}
  	The probability density function (PDF) of the RWP-based Nakagami-$m$ fading channel model is derived as follows:
  	\begin{align}
  		\begin{split}
  		f_{\lVert \boldsymbol{\rm {h}}\rVert^2}(\hspace{-0.5mm}x\hspace{-0.5mm})&\hspace{-0.5mm}=\hspace{-0.5mm}\hspace{-0.5mm}\left[\hspace{-0.5mm}\sum_{n=1}^{N_T} \frac{B_n}{D_U^{\Upsilon_n+1}}\hspace{-0.5mm}\times\hspace{-0.5mm}\gamma\hspace{-0.5mm}\left(\hspace{-0.5mm}\frac{\Upsilon_n\hspace{-1mm}+\hspace{-1mm}1}{\alpha}\hspace{-0.5mm}+\hspace{-0.5mm}N_fm_{N}\hspace{-0.5mm},\hspace{-0.5mm}\frac{m_{N}x}{p_t}D_U^\alpha\hspace{-1mm}\right)\right.\\
  		&\quad\left.\hspace{-0.5mm}-\hspace{-0.5mm}\sum_{n=1}^{N_T} \frac{B_n}{D_L^{\Upsilon_n\hspace{-0.5mm}+\hspace{-0.5mm}1}}\hspace{-1mm}\times\hspace{-1mm}\gamma\hspace{-0.5mm}\left(\hspace{-0.5mm}\frac{\Upsilon_n\hspace{-1mm}+\hspace{-1mm}1}{\alpha}\hspace{-1mm}+\hspace{-1mm}N_fm_{N},\hspace{-1mm}\frac{m_{N}x}{p_t}D_L^\alpha\hspace{-0.5mm}\right)\hspace{-1mm}\right]\\
  		&\quad\times\frac{1}{\alpha\Gamma(N_fm_{N})}\left(\frac{m_{N}}{p_t}\right)^{-\frac{\Upsilon_n+1}{\alpha}}\hspace{-2mm}x^{-\left(\frac{\Upsilon_n+1}{\alpha}+1\right)}\hspace{-0.5mm},
  		\end{split}
  		\label{formula:nakagami-RWP}
  		\end{align}
  	where $\gamma(\cdot)$ denotes the lower incomplete Gamma function. The moving range of the UE is given as $D_L\le r\le D_U $, where $D_L$ and $D_U$ represent maximum and minimum distances between the BS and the UE, respectively. The notations $B_n$, $\Upsilon_n$, and $N_T$ represent  parameters of the RWP model \cite{Cao1}. The notations $m_{N}$ and $\boldsymbol{\rm {h}}$ represent the Nakagami-$m$ fading parameter and the $N_f\times 1$ channel, respectively. The notations $\alpha$ $\Gamma(\cdot)$, and $p_t$, are the path loss exponent, the Gamma function, and the transmit power, respectively.
  \end{theorem}

   \textit{proof:} Please see \cite{Cao1}.$\hfill\blacksquare$.
  
  
   Note that the imperfect CSI of jammers and interferers can be obtained, thus the channel coefficients between the $q$th jammer and the $k$th UE, between the $q$th jammer and the active RIS, and between the $b$th interferer and the $k$th UE, can be modeled, respectively, as $\boldsymbol {\rm h}_{JU,qk}=\widehat{\boldsymbol {\rm h}}_{JU,qk}+\Delta\boldsymbol {\rm h}_{JU,qk}$, $\boldsymbol {\rm G}_{JR,q}=\widehat{\boldsymbol {\rm G}}_{JR,q}+\Delta\boldsymbol {\rm G}_{JR,q}$, and $\boldsymbol {\rm h}_{IU,bk}=\widehat{\boldsymbol {\rm h}}_{IU,bk}+\Delta\boldsymbol {\rm h}_{IU,bk}$,
  where $\widehat{\boldsymbol {\rm h}}_{JU,qk}$, $\widehat{\boldsymbol {\rm G}}_{JU,qk}$, and $\widehat{\boldsymbol {\rm h}}_{IU,bk}$ represent the estimation of the channels between the $q$th jammer and the $k$th UE, between the $q$th jammer and the active RIS, and between the $b$th interferer and the $k$th UE, respectively. In addition, $\Delta\boldsymbol {\rm G}_{JR,q}$, $\Delta\boldsymbol {\rm G}_{JR,q}$, and $\Delta\boldsymbol {\rm h}_{IU,bk}$ refer to the errors of the corresponding channels. Minimum mean square error (MSE) estimation is used, where the  estimation error is uncorrelated with the estimated channel coefficient. According to the above model, the actual channel coefficients is modeled as realizations of $\mathcal{F}\left\{\boldsymbol {\rm h}_{JU,qk}(\vartheta), \boldsymbol {\rm G}_{JR,q}(\vartheta), \boldsymbol {\rm h}_{IU,bk}(\vartheta), \forall\vartheta\right\}$, where $\vartheta$ represents the index of random realizations extracted from the sample space $\mathcal{F}$.
  

  \vspace{-2mm}

  \subsection{Signal Transmission Model}
  \label{subsec:Signal_model}

\begin{figure}[t!]
  	\vspace{-15pt}
  	\centering
  	\includegraphics[scale=0.12]{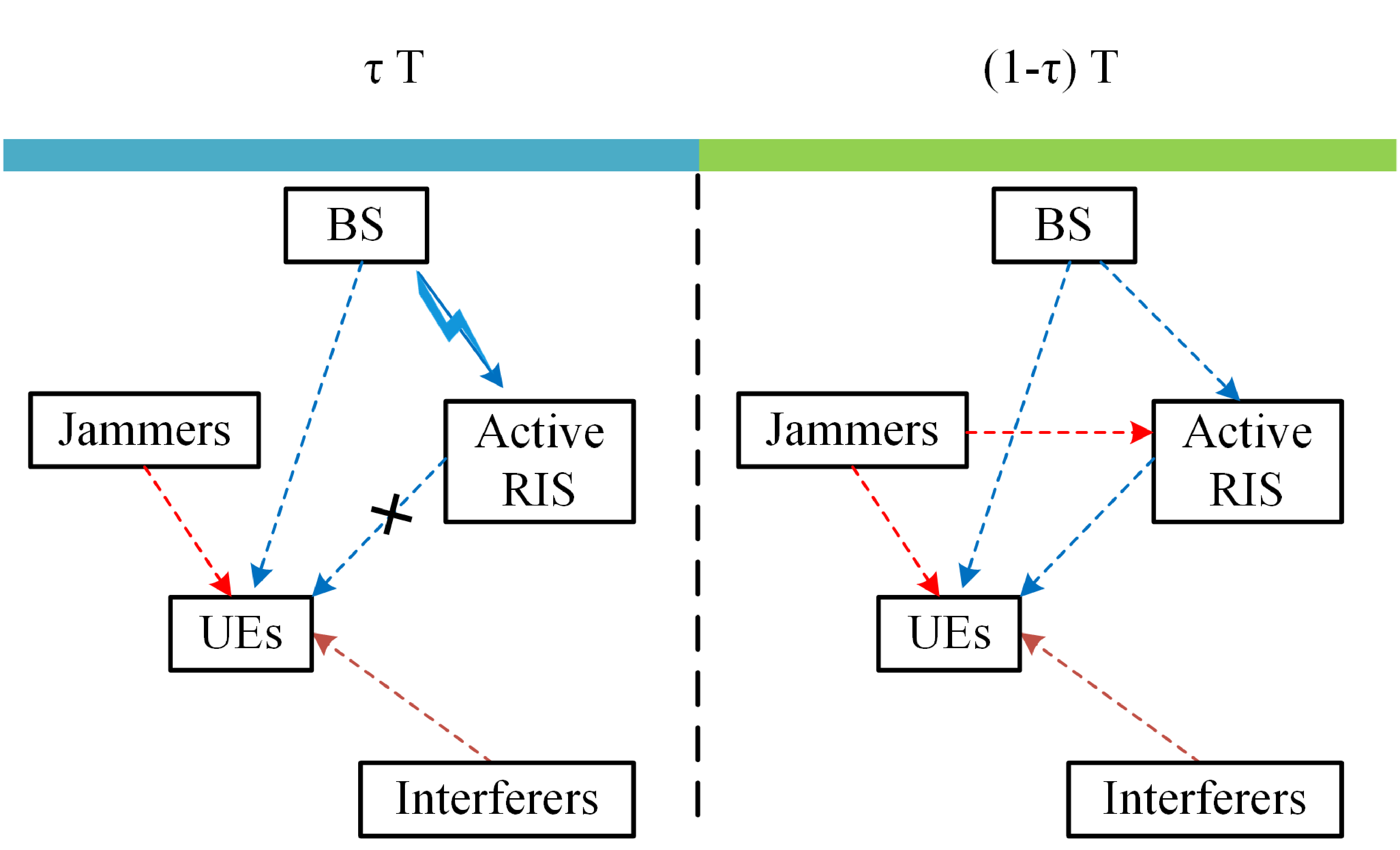}
  	\vspace{-5pt}
  	\caption{The signal transmission stage according to the TD-SWIPT scheme.}
  	\vspace{-15pt}
  	\label{fig:STM-scheme}
  \end{figure}

  Without loss of generality, the communication transmission process is realized in the transmission stage with unit period ($T=1$). As illustrated in Fig.~\ref{fig:STM-scheme}, based on the TD-SWIPT scheme, the transmission stage can be divided into two phases in accordance with the time dividing factor, which is denoted as $\tau$ with $0<\tau<1$. In the first stage (i.e., during the $\tau T$ period), the active RIS is in the state of energy harvesting. The active RIS harvests and stores the energy from the BS transmit signals without transmitting the reflected signal to UEs. In the second stage (i.e., during the $\left(1-\tau\right) T$ period), the active RIS switches to reflection amplification state, and then the BS provides communication services to UEs with the assistance of the active RIS. Besides, in these two phases, jammers can detect the communication activity of the UE thus continuously transmitting malicious jamming signals to interfere with communication. 
  
  In the first stage ($\tau T$ period), based on the TD-SWIPT scheme, the transmitted signal vector, denoted by $\boldsymbol{\rm {x}}_1$, is as follows:
  \begin{align}
  	\boldsymbol{\rm {x}}_1 = \sum_{k=1}^{K}\boldsymbol{\rm {w}}_{1,k}s_{1,k},\quad \forall k\in \mathcal{K},
  \end{align}
  where $\boldsymbol{\rm {w}}_{1,k}\in \mathbb{C}^{N\times 1}$ and $s_{1,k}$ are the transmit beamforming vector and the transmit symbol at the BS in the first stage, respectively.  Besides, $s_{1,k}$ is modeled as $s_{1,k}\sim\mathcal{CN}(0,1)$. Then, the transmit power constraint is denoted as: $\sum_{k=1}^{K}\Vert\boldsymbol{\rm {w}}_{1,k}\Vert^2\le P_{\max}$ with $P_{\max}$ being the maximum transmit power of the BS. The active RIS is energized by extra power source to change the reflected signal. Then, we develop the self-sustainable structure, where the power of transmitted signals is used as the external energy source of the active RIS through TD-SWIPT scheme. Specifically, in the first stage, the harvested power at the active RIS for energy supply, denoted by $E_{R}$, is modeled by
 \begin{equation}
     E_{R}(\boldsymbol{\rm {w}}_{1,k},\tau) = \sum_{k=1}^{K}\tau\eta_1\Vert\boldsymbol{\rm {G}}_{BR}\boldsymbol{\rm {w}}_{1,k}\Vert^2,
 \end{equation}
 where $\eta_1$ denotes the energy harvesting efficiency. In the power harvested of the active RIS, the power harvested from the transmit beamforming at the BS is significantly higher than that of the additive Gaussian white noise, so we neglect the trivial latter.

 The $q$th malicious jammer and the $b$th interferer transmit the jamming signal $\boldsymbol{\rm {z}}_{J,qk}\in\mathbb{C}^{N_{Jam}\times 1} $ and co-channel interference signal $\boldsymbol{\rm {z}}_{I,bk}\in\mathbb{C}^{N\times 1} $ to the $k$th UE when the BS transmits the legitimate signal to the $k$th UE. Hence, the signal received by the $k$th UE is derived as follows:
 	 \begin{align}
 	\begin{split}
 	y_{1,k}=\boldsymbol{\rm h}_{BU,k}^H\boldsymbol{\rm {x}}_1&+\sum_{q=1}^{Q}\boldsymbol{\rm h}_{JU,qk}^H\boldsymbol{\rm z}_{J,qk}s_{J,qk}\\
 	&+\sum_{b=1}^{B}\boldsymbol{\rm h}_{IU,bk}^H\boldsymbol{\rm z}_{I,bk}s_{I,bk}+n_{1,k},
 	\end{split}
 	\end{align}
  where $n_{1,k}\sim\mathcal{CN}(0,\sigma^2_{1})$ represents the additive complex Gaussian noise at the UE in the first stage. Besides, $s_{J,qk}$ and $s_{I,bk}$ denote the transmit symbols at the $q$th jammer and $b$th interferer, respectively. Then, the received SINR is derived as
  \begin{align}
  \begin{split}
  \gamma_{1,k}(\boldsymbol{\rm {w}}_{1,k})=\frac{\left\lvert\boldsymbol{\rm h}_{BU,k}^H\boldsymbol{\rm {w}}_{1,k}\right\rvert^2}{\sum\limits_{j=1,j\neq k}^{K}\left\lvert\boldsymbol{\rm h}_{BU,k}^H\boldsymbol{\rm {w}}_{1,j}\right\rvert^2+Z_{1,k}+\sigma^2_{1}},
  \end{split}
  \end{align}
  where $Z_{1,k}=\sum_{q =1}^{Q}\left\lvert\boldsymbol{\rm h}_{JU,qk}^H\boldsymbol{\rm z}_{J,qk}\right\rvert^2+\sum_{b=1}^{B}\left\lvert\boldsymbol{\rm h}_{IU,bk}^H\boldsymbol{\rm z}_{I,bk}\right\rvert^2$.
  In the first stage, the achievable rate of the $k$th UE can be given as follows:
  \begin{equation}
  R_{1,k}(\boldsymbol{\rm {w}}_{1,k}) = \mbox{log}_2(1+\gamma_{1,k}(\boldsymbol{\rm {w}}_{1,k})).
  \end{equation}

 In the second stage, the active RIS is switched to the reflected information state for reflection and amplification of signals from the BS. The transmitted signal vector, denoted by $\boldsymbol{\rm {x}}_{2}$, at the BS is as follows:
  \begin{align}
  \boldsymbol{\rm {x}}_{2}= \sum_{k=1}^{K}\boldsymbol{\rm {w}}_{2,k}s_{2,k},\quad \forall k\in \mathcal{K},
  \end{align}
  where $\boldsymbol{\rm {w}}_{2,k}\in \mathbb{C}^{N\times 1}$ and $s_{2,k}$ are the transmit beamforming vector at the $b$th BS and the transmit symbol in the second stage, respectively. The transmit symbol $s_{2,k}$ is modeled as $s_{2,k}\sim\mathcal{CN}(0,1)$. Similar to the first stage, we consider the constraint: $\sum_{k=1}^{K}\Vert\boldsymbol{\rm {w}}_{2,k}\Vert^2\le P_{\max}$. Hence, the signal received by the $k$th UE is derived as follows:
  \begin{align}\label{formula:received_y}
  		y_{2,k}\hspace{-1mm}=&\boldsymbol{\rm {h}}_{BU,k}^H\boldsymbol{\rm {x}}_{2}\hspace{-1mm}+\hspace{-1mm}\boldsymbol{\rm {h}}_{RU,k}^H\boldsymbol{\rm {\Theta}}\hspace{-1mm}\left(\boldsymbol{\rm {G}}_{BR}\boldsymbol{\rm {x}}_{2}\hspace{-1mm}+\hspace{-1mm}\boldsymbol{\rm {n}}_{R}\right)\hspace{-1mm}+\hspace{-1.5mm}\sum_{b=1}^{B}\boldsymbol{\rm {h}}_{IU,bk}^H\boldsymbol{\rm {z}}_{I,bk}s_{I,bk}\nonumber\\
  		&\hspace{-1mm}+\hspace{-1mm}\sum_{q=1}^{Q}\left(\boldsymbol{\rm {h}}_{JU,qk}^H\hspace{-1mm}+\hspace{-1mm}\boldsymbol{\rm {h}}_{RU,k}^H\boldsymbol{\rm {\Theta}}\boldsymbol{\rm {G}}_{JR,q}^H\right)\boldsymbol{\rm {z}}_{J,qk}s_{J,qk}\hspace{-1mm}+\hspace{-1mm}n_{2,k},
  \end{align}
 where $n_{2,k}\sim\mathcal{CN}(0,\sigma^2_{2})$ and $\boldsymbol{n}_{R}\sim\mathcal{CN}(0,\sigma^2_R\boldsymbol{\rm I}_M)$ represent the additive complex Gaussian noise at the $k$th UE and the active RIS, respectively. The received signals of the $k$th UE are respectively composed of the legitimate signals from the BS, malicious jamming signals from jammers, co-channel interference signals from interferers, and channel noise in the communication system.
 
 For simplicity, we define $\boldsymbol{\rm h}_{k}^H=\boldsymbol{\rm h}_{BU,k}^H+\boldsymbol{\rm h}_{RU,k}^H\boldsymbol {\rm \Theta}\boldsymbol{\rm G}_{BR}$ and $\boldsymbol{\rm h}_{J,qk}^H=\boldsymbol{\rm h}_{JU,qk}^H+\boldsymbol{\rm h}_{RU,k}^H\boldsymbol {\rm \Theta}\boldsymbol{\rm G}_{JR,q}$. Then, the received SINR is written as follows:
 \begin{align}
 	\begin{split}
 		&\gamma_{2,k}(\boldsymbol{\rm {w}}_{2,k},\boldsymbol{\rm {\Theta}})\\
 		&=\frac{\left\lvert\boldsymbol{\rm h}_{k}^H\boldsymbol{\rm {w}}_{1,k}\right\rvert^2}{\sum\limits_{j=1,j\neq k}^{K}\left\lvert\boldsymbol{\rm h}_{k}^H\boldsymbol{\rm {w}}_{1,j}\right\rvert^2+\sigma^2_{R}\left\Vert\boldsymbol{\rm {h}}_{RU,k}^H\boldsymbol{\rm {\Theta}}\right\Vert^2+Z_{2,k}+\sigma^2_{2}},
 	\end{split}
 \end{align}
 where $Z_{2,k}=\sum_{q =1}^{Q}\left\lvert\boldsymbol{\rm h}_{J,qk}^H\boldsymbol{\rm z}_{J,qk}\right\rvert^2+\sum_{b=1}^{B}\left\lvert\boldsymbol{\rm h}_{IU,b}^H\boldsymbol{\rm z}_{I,bk}\right\rvert^2$. The achievable rate of UE in the second stage is given as follows:
 \begin{equation}
 R_{2,k}(\boldsymbol{\rm {w}}_{2,k},\boldsymbol{\rm {\Theta}}) = \mbox{log}_2(1+\gamma_{2,k}(\boldsymbol{\rm {w}}_{2,k},\boldsymbol{\rm {\Theta}})).
 \end{equation}
 
 Based on the above-mentioned TD-SWIPT scheme, the achievable rate in unit period ($T=1$) can be obtained as follows:
 \begin{align}
 \begin{split}
 R_{\rm{sum}}&(\tau,\boldsymbol{\rm {w}}_{1,k},\boldsymbol{\rm {w}}_{2,k},\boldsymbol{\rm {\Theta}}) \\
 &=\sum_{k=1}^{K}\left(\tau R_{1,k}(\boldsymbol{\rm {w}}_{1,k})+\left(1-\tau\right)R_{2,k}(\boldsymbol{\rm {w}}_{2,k},\boldsymbol{\rm {\Theta}})\right).
 \end{split}
 \end{align}
\vspace{-7mm}
\section{Problem Formulation}
\label{sec:Problem_formulation}

\subsection{Power Consumption Model}
\label{subsec:Power_Consumption} 
 Additional power consumption is caused by the introduction of the active RIS. By using TD-SWIPT scheme, the self-sustainable active RIS is powered by harvesting energy from the signal transmitted by the BS. In comparison to the passive RIS, each RE of the active RIS is added a reflection amplifier (RA), which can be implemented by a tunnel diode circuit. Then, we adopt the conventional power amplifier model \cite{Energy-Efficient} to characterize the power consumption of the $m$th RE, denoted by $P_{RE,m}$, as follows:
 \begin{equation}
 P_{RE,m} = P_{dc}+P_{sc}+\xi P_{o,m},
 \end{equation}
 where $P_{o,m}$, $P_{dc}$, $P_{sc}$, and $\xi$ are the output power of the $m$th RE, the direct current (DC) biasing power, the control circuit power, and the reciprocal of the amplification efficiency, respectively. In addition, the output power, denoted by $P_{o,m}$, at the $m$th RE is derived as follows:
\begin{equation}\vspace{-1mm}
 P_{o,m} = A^2_mP_{i,m},
\end{equation}
 where $P_{i,m}$ and $A_m$ represent the incident signal power and the amplitude of the $m$th RE at the active RIS, respectively. Practically, the incident signals arriving at the active RIS are weakened by attenuation, which enables the active RIS to obtain superior amplification yield with a lower $P_{o,m}$ \cite{Active-Reconfigurable}. The RE of the active RIS can amplify signals in a low energy consumption way with an simple structure. 
  

 In the second stage, the incident signal power at active RIS with $M$ REs for information transmission is described as follows:
  \begin{equation}
  P_{I} = \sum_{k=1}^{K}\Vert\boldsymbol{\rm {G}}_{BR}\boldsymbol{\rm {w}}_{2,k}\Vert^2+\sigma^2_R.
  \end{equation}
  Then, the output power of active RIS is expressed as follows:
  \begin{equation}\vspace{-2mm}
  P_{O} = \sum_{k=1}^{K}\Vert\boldsymbol {\rm \Theta}\boldsymbol{\rm {G}}_{BR}\boldsymbol{\rm {w}}_{2,k}\Vert^2+\sigma^2_R\Vert\boldsymbol {\rm \Theta}\boldsymbol {\rm I}_M\Vert^2.
  \end{equation}
  Hence, the total power consumption is further written as follows:
  \begin{align}
  P_{R}\left(\boldsymbol{\rm {w}}_{2,k},\boldsymbol {\rm \Theta}\right)=&\xi\hspace{-1mm} \left(\sum_{k=1}^{K}\hspace{-1mm}\Vert\boldsymbol {\rm \Theta}\boldsymbol{\rm {G}}_{BR}\boldsymbol{\rm {w}}_{2,k}\Vert^2\hspace{-1mm}+\hspace{-1mm}\sigma^2_R\Vert\boldsymbol {\rm \Theta}\boldsymbol {\rm I}_M\Vert^2\hspace{-1mm}\right) \hspace{-1mm}\nonumber\\ 
  &\qquad\qquad\qquad+\hspace{-1mm} M(P_{dc}+P_{sc}).
  \end{align}
  
   Moreover, we consider that the active RIS is supplied by the energy obtained from the BS signals , so the energy supply constraint is derived as follows:
  \begin{equation}
  E_{R}(\boldsymbol{\rm {w}}_{1,k},\tau) \ge \left(1-\tau\right)P_{R}\left(\boldsymbol{\rm {w}}_{2,k},\boldsymbol {\rm \Theta}\right).
  \end{equation}

  The configuration of the active RIS entails the provision of a RA associated with each RE. Thus, the active RIS empowers REs with the capability to independently change the incoming signal. REs at the active RIS achieve lightweight amplification of signals without complex and power-consuming RF chain. However, the RAs with low power consumption limit the amplification of reflected signals. Therefore, we set the maximum amplitude limited by the working principle of the RA as $A_{\max}$. For each RE, the amplitude constraint is presented as follows:
  \begin{equation}
  A_{m} \le A_{\max},\forall m \in \mathcal{M}. 
  \end{equation}

\subsection{Problem Formulation}
\label{subsec:Problem_formulation} 

  We develop an optimization problem of maximizing the achievable rate of the anti-jamming system by jointly designing the time dividing factor $\tau$, the transmit beamforming $\boldsymbol{\rm {w}}_{1,k}$ and $\boldsymbol{\rm {w}}_{2,k}$, and the reflecting beamforming matrix $\boldsymbol{\rm {\Theta}}$, subject to energy supply, amplification power, maximum transmit power, and maximum amplification amplitude. Specifically, based on the above imperfect CSI settings, the maximization problem $\textbf{\textit{P}1}$ is derived as follows:
 \begin{subequations}\label{Problem:yhwt1}
	\begin{alignat}{2}
		\textbf{\textit{P}1:}
		&\mathop{\max}\limits_{\tau,\boldsymbol{\rm {w}}_{1,k},\boldsymbol{\rm {w}}_{2,k},\boldsymbol{\rm {\Theta}}} \mathbb{E}_{\vartheta}\left[R_{\rm{sum}}(\tau,\boldsymbol{\rm {w}}_{1,k},\boldsymbol{\rm {w}}_{2,k},\boldsymbol{\rm {\Theta}};\vartheta)\right]  \notag \\
		{\rm{s.t.}}:&1).\ \sum_{k=1}^{K}\Vert\boldsymbol{\rm {w}}_{1,k}\Vert^2\le P_{\max}, \sum_{k=1}^{K}\Vert\boldsymbol{\rm {w}}_{2,k}\Vert^2\le P_{\max};\\
		&2).\ E_{R}(\boldsymbol{\rm {w}}_{1,k},\tau) \ge \left(1-\tau\right)P_{R}\left(\boldsymbol{\rm {w}}_{2,k},\boldsymbol {\rm \Theta}\right); \\
		&3).\  A_{m} \le A_{\max},\forall m \in \mathcal{M}.
	\end{alignat}
 \end{subequations} 

  Noting that $\textbf{\textit{P}1}$ is a nonconvex optimization problem owing to severe coupled optimization variables and the fractional form of objective function of $\textbf{\textit{P}1}$. Furthermore, $\textbf{\textit{P}1}$ is constructed as a stochastic optimization problem by introducing expectations, which further complicates the problem.

 \section{Alternating Optimization}
 \label{sec:AO}

 The main challenge in solving $\textbf{\textit{P}1}$ originates from the stochastic nature of the objective and constraint functions, which introduces expectation values hard to compute exactly. In the special case, where $\vartheta$ is a deterministic number, $\textbf{\textit{P}1}$ degenerates into a deterministic optimization problem, i.e., an optimization problem in the perfect CSI case. However, in the stochastic case, it is considerably more challenging to design an algorithm to solve problems involving the stochastic nonconvex objective function.
 
 The classical sample average approximation (SAA) method is used to deal with $\textbf{\textit{P}1}$. Thus, at the $r$th iteration, $\textbf{\textit{P}1}$ is redescribed as follows:
  \begin{subequations}\label{Problem:yhwt1A}
 	\begin{alignat}{2}
 		\textbf{\textit{P}1-A:}
 		&\mathop{\max}\limits_{\tau,\boldsymbol{\rm {w}}_{1,k},\boldsymbol{\rm {w}}_{2,k},\boldsymbol{\rm {\Theta}}}\frac{1}{r}\sum_{i=1}^{r} R_{\rm{sum}}(\tau,\boldsymbol{\rm {w}}_{1,k},\boldsymbol{\rm {w}}_{2,k},\boldsymbol{\rm {\Theta}};\vartheta_i) \notag \\
 		{\rm{s.t.}}:& (\ref{Problem:yhwt1}{\text{a}}),(\ref{Problem:yhwt1}{\text{b}}),~\text{and}~ (\ref{Problem:yhwt1}{\text{c}})\nonumber.
 	\end{alignat}
 \end{subequations} 
 However, the SAA computational complexity is high, because the average of numerous random samples is computed for each iteration. In addition, some of the variables to be optimized are highly coupled. To overcome the above mentioned difficulties, we adopt the SSCA-based AO algorithm, where the SSCA scheme devises suitable convex approximation surrogate function for the objective function at each iteration and at each channel realization, thereby determining the solution at the $r$th iteration. Thus, based on the SSCA-based AO algorithm, the $\textbf{\textit{P}1-A}$ is discretized as four subproblems to update the corresponding variables.

 \subsection{Optimize $\tau$ And $\boldsymbol{\rm {w}}_{1,k}$}
 \label{subsec:a1}
 
 \subsubsection{Optimize the time dividing factor $\tau$}
 We decompose $\textbf{\textit{P}1-A}$ into several subproblems by exploiting the structural properties of $\textbf{\textit{P}1-A}$. To begin with, we optimize the time dividing factor $\tau$, with the other variables fixed. In the objective function, we note that $\boldsymbol{\rm {w}}_{2,k}$ and $\boldsymbol{\rm {\Theta}}$ appear only in $R_{2,k}$ and $\boldsymbol{\rm {w}}_{1,k}$ appears only in $R_{1,k}$. Thus, when $\boldsymbol{\rm {w}}_{1,k}$, $\boldsymbol{\rm {w}}_{2,k}$, and $\boldsymbol{\rm {\Theta}}$ are given, $R_{1,k}$ and $R_{2,k}$ correspond to the given and are independent of the value of $\tau$. Let the stationary values of $R_{1,k}$ and $R_{2,k}$  with $\boldsymbol{\rm {w}}_{1,k}$, $\boldsymbol{\rm {w}}_{2,k}$, and $\boldsymbol{\rm {\Theta}}$ fixed at $i$th channel realization be $\bar{R}_{1,k}(\vartheta_i)$ and $\bar{R}_{2,k}(\vartheta_i)$, respectively. Thus, the subproblem for $\tau$ is given as follows:
 \begin{subequations}\label{Problem:yhwt1t}
 	\begin{alignat}{2}
 		\textbf{\textit{P}2-A:}
 		&\mathop{\max}\limits_{\tau}\frac{1}{r}\sum_{i=1}^{r} \sum_{k=1}^{K}\left(\tau \bar{R}_{1,k}(\vartheta_i)+\left(1-\tau\right)\bar{R}_{2,k}(\vartheta_i)\right) \notag \\
 		{\rm{s.t.}}:&1). \ E_{R}(\tau) \ge \left(1-\tau\right)P_{R}.
 	\end{alignat}
 \end{subequations} 
 Note that owing to channels are enhanced by the active RIS in the second stage, $\bar{R}_{1,k}(\vartheta_i)\le \bar{R}_{2,k}(\vartheta_i)$ is satisfied. Thus, we can easily verify that the constraint~(\ref{Problem:yhwt1t}{a}) is equivalent at the optimum of subproblem $\textbf{\textit{P}2-A}$. Thus, the optimal $\tau^{\star}$ is deduced as follows:
\begin{align}\label{tau}
	\tau^{\star}=\frac{P_{R}}{P_{R}+\sum\limits_{k=1}^{K}\eta_1\Vert\boldsymbol{\rm {G}}_{BR}\boldsymbol{\rm {w}}_{1,k}\Vert^2}.
\end{align}

\subsubsection{Optimize the transmit beamforming vector $\boldsymbol{\rm {w}}_{1,k}$}

 Then, we optimize the first-stage transmit beamforming $\boldsymbol{\rm {w}}_{1,k}$ with the given $\boldsymbol{\rm {w}}_{1,k}$, $\boldsymbol{\rm {\Theta}}$, and $\tau$. 
 The objective function is a fractional form, which requires further processing. Furthermore, based on the Quadratic Transform, the objective function of $\textbf{\textit{P}1-A}$ is transformed by introducing auxiliary variables $\boldsymbol{\rm {\omega}}_1 =\left[\omega_{1,1},\cdots,\omega_{1,K}\right]^T$ and $\boldsymbol{\rm {\nu}}_1=\left[\nu_{1,1},\cdots,\nu_{1,K}\right]^T$ and omitting irrelevant items as follows:
 	\begin{align}
 	&\hspace{-2mm}f_{OF}^I\left(\boldsymbol{\rm {w}}_{1,k},\boldsymbol{\rm {\omega}}_1,\boldsymbol{\rm {\nu}}_1\right)\nonumber\\
 	=&\frac{1}{r}\hspace{-1mm}\sum_{i=1}^{r}\hspace{-1mm}\sum_{k=1}^{K}\hspace{-1mm}\left[\mbox{ln}(1\hspace{-1mm}+\hspace{-1mm}\omega_{1,k})\hspace{-1mm}+\hspace{-1mm}2\sqrt{1\hspace{-1mm}+\hspace{-1mm}\omega_{1,k}}\mbox{Re}\hspace{-1mm}\left\{\hspace{-1mm}\nu_{1,k}^*\boldsymbol{\rm {h}}_{BU,k}^H\boldsymbol{\rm {w}}_{1,k}\hspace{-1mm}\right\}\right.\nonumber\\
 	&-\hspace{-1mm}\omega_{1,k}\hspace{-1mm}-\hspace{-1mm}\left.\left\vert\nu_{1,k}\right\vert^2\hspace{-1mm}\left(\hspace{-1mm}\textstyle\sum\limits_{j=1}^{K}\left\vert\boldsymbol{\rm {h}}_{BU,k}^H\boldsymbol{\rm {w}}_{1,j}\right\vert^2\hspace{-2mm}+\hspace{-1mm}Z_{1,k}(\vartheta_i)+\hspace{-1mm}\sigma^2_{1}\hspace{-1mm}\right)\right]\hspace{-1mm},
 	\end{align}
 where $Z_{1,k}(\vartheta_i)$ represents $Z_{1,k}$ at $i$th channel realization. Thus, $\textbf{\textit{P}1-A}$ is reconstructed as follows:
 
  \begin{subequations}\label{Problem:yhwt1-1}
 	\begin{alignat}{2}
 		\textbf{\textit{P}2-B:}
 		&\mathop{\max}\limits_{\boldsymbol{\rm {w}}_{1,k},\boldsymbol{\rm {\omega}}_1,\boldsymbol{\rm {\nu}}_1} f_{OF}^{I}\left(\boldsymbol{\rm {w}}_{1,k},\boldsymbol{\rm {\omega}}_1,\boldsymbol{\rm {\nu}}_1\right)\nonumber\\
 		{\rm{s.t.}}:&1).\ \sum_{k=1}^{K}\Vert\boldsymbol{\rm {w}}_{1,k}\Vert^2\le P_{\max};\\
 		&2).\ E_{R}(\boldsymbol{\rm {w}}_{1,k},\tau) \ge \left(1-\tau\right)P_{R}.
 	\end{alignat}
 \end{subequations} 
 Note that for all $k\in \mathcal{K}$, by letting $\partial f_{OF}^I/\partial \omega_{1,k}$ and $\partial f_{OF}^I/\partial\nu_{1,k}$ be zero, the optimal $\omega_{1,k}^{\star}$ and $\nu_{1,k}^{\star}$ are obtained, respectively, as follows:	\begin{align}\label{omage_k}
 	\omega_{1,k}^{\star}=&\frac{\left\lvert\boldsymbol{\rm h}_{BU,k}^H\boldsymbol{\rm {w}}_{1,k}\right\rvert^2}{\sum\limits_{j=1,j\neq k}^{K}\left\lvert\boldsymbol{\rm h}_{BU,k}^H\boldsymbol{\rm {w}}_{1,j}\right\rvert^2+\frac{1}{r}\hspace{-1mm}\sum\limits_{i=1}^{r}\hspace{-1mm}Z_{1,k}(\vartheta_i)+\sigma^2_{1}},\\
 	\nu_{1,k}^{\star}=&\frac{\sqrt{1-\omega_{1,k}}\boldsymbol{\rm h}_{BU,k}^H\boldsymbol{\rm {w}}_{1,k}}{\sum\limits_{j=1}^{K}\hspace{-1mm}\left\vert\boldsymbol{\rm h}_{BU,k}^H\boldsymbol{\rm {w}}_{1,j}\right\vert^2\hspace{-2mm}+\frac{1}{r}\hspace{-1mm}\sum\limits_{i=1}^{r}\hspace{-1mm}Z_{1,k}(\vartheta_i)+\sigma^2_{1}}.\label{nu_k}
 	\end{align}

  However, owing to the non-convexity of the constraint  (\ref{Problem:yhwt1-1}{b}), $\textbf{\textit{P}2-B}$ is still the non-convex problem. Therefore, we consider dealing with this constraint effectively by SCA method. By applying Jensen' inequality, we approximate it, as follow:
  \begin{align}\label{formula:new_opti_1}
  \begin{split}
  \Vert\boldsymbol{\rm {G}}_{BR}\boldsymbol{\rm {w}}_{1,k}\Vert^2=&\boldsymbol{\rm {w}}_{1,k}^H\boldsymbol{\rm {G}}_{BR}^H\boldsymbol{\rm {G}}_{BR}\boldsymbol{\rm {w}}_{1,k}\\
  \ge&2\mbox{Re}\left\{\boldsymbol{\rm {w}}_{1,k}^{\left(i\right)H}\boldsymbol{\rm {K}}_{1}\left(\boldsymbol{\rm {w}}_{1,k}-\boldsymbol{\rm {w}}_{1,k}^{\left(i\right)}\right)\right\}\\
  &+\boldsymbol{\rm {w}}_{1,k}^{\left(i\right)H}\boldsymbol{\rm {K}}_{1}\boldsymbol{\rm {w}}_{1,k},
  \end{split}
  \end{align}
  where $\boldsymbol{\rm K}_{1}=\boldsymbol{\rm {G}}_{BR}^H\boldsymbol{\rm {G}}_{BR}$, and $\boldsymbol{\rm {w}}_{1,k}^{\left(i\right)}$ indicates the transmit beamforming vector in the $i$th iteration. By substituting (\ref{formula:new_opti_1}), the constraint (\ref{Problem:yhwt1-1}{b}) can be rewritten as follows:
  \begin{align}\label{formula:new_opti_2}
  \begin{split}
  2\tau\eta_1\sum_{k=1}^{K}\mbox{Re}\left\{\boldsymbol{\rm {w}}_{1,k}^{\left(i\right)H}\boldsymbol{\rm K}_{1}\boldsymbol{\rm {w}}_{1,k}\right\}\ge\Xi_{1},
  \end{split}
  \end{align}
  where $\Xi_{1}=\left(1-\tau\right)P_{R}+\tau\eta_1\sum_{k=1}^{K}\boldsymbol{\rm {w}}_{1,k}^{\left(i\right)H}\boldsymbol{\rm K}_{1}\boldsymbol{\rm {w}}_{1,k}^{\left(i\right)}$. Thus, upon replacing the constraint (\ref{Problem:yhwt1-1}{b}) by (\ref{formula:new_opti_2}), we have the subproblem with respect to $\boldsymbol{\rm {w}}_{1,k}$ as follows:
   \begin{subequations}\label{Problem:yhwt_new11}
  	\begin{alignat}{2}
  	\textbf{\textit{P}2-C:}
  	\mathop{\min}\limits_{\boldsymbol{\rm {w}}_{1,k}}f_{OF}^{I}\left(\boldsymbol{\rm {w}}_{1,k}\right)\quad {\rm{s.t.}}: (\ref{Problem:yhwt1-1}{\text{a}})\ \text{and}\ (\ref{formula:new_opti_2}),\notag
  	\end{alignat}
   \end{subequations}
   which constitutes a convex optimization problem. Then, we employ the Lagrangian dual decomposition method to solve the approximate optimal solutions with low complexity. Note that the dual gap is zero since $\textbf{\textit{P}2-C}$ is a convex problem satisfying slater's condition \cite{PAN-CUN}. To begin with, by introducing the Lagrange multiplier $\lambda_{1}$ and $\lambda_{2}$ associated with the constraint (\ref{Problem:yhwt1-1}{a}) and (\ref{Problem:yhwt1-1}{b}) and omitting irrelevant items, we can derive the Lagrangian function as follows:

  \begin{small}\vspace{-3mm}
  	\begin{align}\label{formula:new_opti_6}
  	\mathcal{L}_1&\hspace{-1mm}\left(\boldsymbol{\rm {w}}_{1,k},\lambda_{1},\lambda_{2}\right)\nonumber\\
  	=&\sum_{k=1}^{K}\hspace{-1mm}2\sqrt{1\hspace{-1mm}+\hspace{-1mm}\omega_{1,k}}\mbox{Re}\hspace{-1mm}\left\{\hspace{-1mm}\nu_{1,k}^*\boldsymbol{\rm {h}}_{BU,k}^H\boldsymbol{\rm {w}}_{1,k}\hspace{-1mm}\right\}\hspace{-1mm}+\hspace{-1mm}\lambda_{1}\hspace{-1mm}\sum\limits_{k=1}^{K}\hspace{-1mm}\Vert\boldsymbol{\rm {w}}_{1,k}\Vert^2\hspace{-1mm}-\hspace{-1mm}\lambda_{1}P_{\max}\nonumber\\
  	&-\sum_{k=1}^{K}\hspace{-1mm}\left\vert\nu_{1,k}\right\vert^2\hspace{-1mm}\left(\hspace{-1mm}\textstyle\sum\limits_{j=1}^{K}\left\vert\boldsymbol{\rm {h}}_{BU,k}^H\boldsymbol{\rm {w}}_{1,j}\right\vert^2\hspace{-2mm}+\frac{1}{r}\hspace{-1mm}\sum\limits_{i=1}^{r}\hspace{-1mm}Z_{1,k}(\vartheta_i)+\hspace{-1mm}\sigma^2_{k}\hspace{-1mm}\right)\hspace{-1mm}\nonumber\\
  	&-2\lambda_{2}\tau\eta_1\sum_{k=1}^{K}\mbox{Re}\left\{\boldsymbol{\rm {w}}_{1,k}^{\left(i\right)H}\boldsymbol{\rm K}_{1}\boldsymbol{\rm {w}}_{1,k}\right\}+\lambda_{2}\Xi_{1}.
  	\end{align}
  \end{small}
  
  The optimal solution of $\boldsymbol{\rm {w}}_{1,k}$ is acquired as follows:
  \begin{align}\label{formula:new_opti_7}
  \begin{split}
  \boldsymbol{\rm {w}}_{1,k}^{\star}\hspace{-1mm}\left(\lambda_{2}\right)=&\left(\boldsymbol{\rm K}_{2,k}\hspace{-1mm}-\hspace{-1mm}\lambda_{1}\boldsymbol{\rm I}_{N}\right)^{-1}\hspace{-1mm}
  \left(\boldsymbol{\rm K}_{3,k}-\lambda_{2}\tau\eta_1\boldsymbol{\rm K}_{1}\boldsymbol{\rm {w}}_{1,k}^{\left(i\right)}\right)\hspace{-1mm},
  \end{split}
  \end{align}
  where $\boldsymbol{\rm K}_{2,k}\hspace{-1mm}=\hspace{-1mm}\left\vert\nu_{1,k}\right\vert^2\boldsymbol{\rm h}_{BU,k}^H\boldsymbol{\rm h}_{BU,k}$ and $\boldsymbol{\rm K}_{3,k}\hspace{-1mm}=\hspace{-1mm}\sqrt{1\hspace{-1mm}+\hspace{-1mm}\omega_{1,k}}\nu_{1,k}\boldsymbol{\rm {h}}_{BU,k}$. When the value of $\lambda_{2}$ is selected, it is inevitable to ensure that the complementary slackness condition of (\ref{formula:new_opti_2}) is fulfilled, as follows:
  \begin{align}\label{formula:new_opti_8}
  \begin{split}
  \lambda_{2}\left( 2\tau\eta_1\sum_{k=1}^{K}\mbox{Re}\left\{\boldsymbol{\rm {w}}_{1,k}^{\left(i\right)H}\boldsymbol{\rm K}_{1}\boldsymbol{\rm {w}}_{1,k}\right\}-\Xi_{1}\right)=0.
  \end{split}
  \end{align}
  Therefore, if the following condition is satisfied:
  \begin{align}\label{alg:yinyong1}
  \begin{split}
   2\tau\eta_1\sum_{k=1}^{K}\mbox{Re}\left\{\boldsymbol{\rm {w}}_{1,k}^{\left(i\right)H}\boldsymbol{\rm K}_{1}\boldsymbol{\rm {w}}_{1,k}\right\}=\Xi_{1},
  \end{split}
  \end{align}
  the optimal $\lambda_{2}^{\star}$ is given as follows:
  \begin{align}\label{alg:yinyong2}
  \begin{split}
  \lambda_{2}^{\star}\hspace{-1mm}=\hspace{-1mm}\frac{\widehat{\Xi}_{1}}{2\hspace{-0.5mm}\left(\hspace{-0.5mm}\tau\eta_1\hspace{-0.5mm}\right)^2\sum\limits_{k=1}^{K}\left(\boldsymbol{\rm {w}}_{1,k}^{\left(i\right)H}\hspace{-0.5mm}\boldsymbol{\rm K}_{1}\hspace{-0.5mm}\left(\hspace{-0.5mm}\boldsymbol{\rm K}_{2,k}\hspace{-0.5mm}-\hspace{-0.5mm}\lambda_{1}\boldsymbol{\rm I}_{N}\hspace{-0.5mm}\right)\hspace{-0.5mm}^{-1}\boldsymbol{\rm K}_{1}\boldsymbol{\rm {w}}_{1,k}^{\left(i\right)}\right)}\hspace{-0.5mm},
  \end{split}
  \end{align}  
 where $\widehat{\Xi}_{1}\hspace{-0.5mm}=\hspace{-0.5mm}2\tau\eta_1\sum_{k=1}^{K}\mbox{Re}\left\{\boldsymbol{\rm {w}}_{1,k}^{\left(i\right)H}\hspace{-0.5mm}\boldsymbol{\rm K}_{1}\hspace{-0.5mm}\left(\hspace{-0.5mm}\boldsymbol{\rm K}_{2,k}\hspace{-1mm}-\hspace{-1mm}\lambda_{1}\boldsymbol{\rm I}_{N}\hspace{-0.5mm}\right)\hspace{-0.5mm}^{-1}\boldsymbol{\rm K}_{3,k}\right\}\hspace{-0.5mm}-\hspace{-0.5mm}\Xi_1.$
  Otherwise, the following condition is satisfied
  \begin{align}\label{alg:yinyong3}
  \begin{split}
  2\tau\eta_1\sum_{k=1}^{K}\mbox{Re}\left\{\boldsymbol{\rm {w}}_{1,k}^{\left(i\right)H}\boldsymbol{\rm K}_{1}\boldsymbol{\rm {w}}_{1,k}\right\}\ge\Xi_{1},
  \end{split}
  \end{align}
  where the optimal $\lambda_{2}$ is equivalent to zero. 
  
  Then, for the given $\lambda_{1}$, the complementary slackness condition of constraint (\ref{Problem:yhwt1-1}{a}) is satisfied, as follows:
  \begin{align}
  \begin{split}
  \lambda_{1}\left(\sum_{k=1}^{K}\Vert\boldsymbol{\rm {w}}_{1,k}\left(\lambda_{1}\right)\Vert^2-P_{\max}\right)=0.
  \end{split}
  \end{align} 
  If the following condition is satisfied, as follows:  
   \begin{align}
  \begin{split}
  \mathcal{P}\left(\lambda_{1}\right)\triangleq\sum_{k=1}^{K}\Vert\boldsymbol{\rm {w}}_{1,k}\left(\lambda_{1}\right)\Vert^2= P_{\max},
  \end{split}
  \end{align} 
  Owing to the complex structure of $\lambda_{2}^{\star}$, we are hard pressed to get the closed expression for $\lambda_{1}$. However, $\mathcal{P}\left(\lambda_{1}\right)$ is proved to be a monotonic decreasing function of $\lambda_{1}$ in \cite[Lemma~1]{PAN-CUN}, allowing the bisection search method to attain $\lambda_{1}$. Otherwise, the optimal $\lambda_{1}$ is $\lambda_{1}^{\star}=0$.
  
   In Algorithm~\ref{alg:SCA_erfen}, specific steps for solving the SCA algorithm for problem $\textbf{\textit{P}2-C}$. In each iteration, the complexity is embodied in the calculation of $\boldsymbol{\rm {w}}_{1,k}^{(i+1)}, \forall k \in \mathcal{K}$ according to the bisection search method \cite{PAN-CUN}. The total complexity of the calculation of $\boldsymbol{\rm {w}}_{1,k}^{(i+1)}, \forall k \in \mathcal{K}$ is $\mathcal{O}\left(N^3\mbox{log}_2\frac{\lambda_{U}-\lambda_{L}}{2}\right)$, where $\lambda_{U}$ and $\lambda_{L}$ represent the upper and lower bounds $\lambda_{1,k}$ in bisection search method. 
  
 \begin{algorithm}[t]
  	 \caption{ SCA Algorithm for Problem \textbf{\textit{P}2-C}.}
  	\label{alg:SCA_erfen}
  	\begin{algorithmic}[1]
  		\Require
  		Initialize the transmit beamformings $\boldsymbol{\rm {w}}_{1,k}^{(0)}$, the number of iterations $i=0$, the accuracy $\varsigma_1=10^{-3}$, and the maximum iterations $i_{max}=15$.
  		\Ensure
  		\State Compute the objective function of $\textbf{\textit{P}2-C}$, expressed by $V_{al,1}\left(\boldsymbol{\rm {w}}_{1,k}\right)$.
  		\State Calculate $\Xi_{1}^{\left(i\right)}=\left(1-\tau\right)P_{R}+\tau\eta_1\sum_{k=1}^{K}\boldsymbol{\rm {w}}_{1,k}^{\left(i\right)H}\boldsymbol{\rm K}_{1}\boldsymbol{\rm {w}}_{1,k}^{\left(i\right)}$.
  		\State With given $\Xi_{1}^{\left(k_1\right)}$, tackle the subproblem $\textbf{\textit{P}2-C}$ and get the optimal solution $\boldsymbol{\rm {w}}_{1,k}^{(i+1)}, \forall k \in \mathcal{K}$ according to the bisection search method, where details of the bisection search algorithm can be found in \cite{PAN-CUN}.
  		\State Update $i=i+1$.
  		\renewcommand{\algorithmicrequire}{\textbf{until}}
  		\Require $\left\vert V_{al,1}\left(\boldsymbol{\rm {w}}_{1,k}^{(i+1)}\right)-V_{al,1}\left(\boldsymbol{\rm {w}}_{1,k}^{(i)}\right)\right\vert/V_{al,1}\left(\boldsymbol{\rm {w}}_{1,k}^{(i+1)}\right)<\varsigma_1$ or $i = i_{\max}$.
  	\end{algorithmic}
  \end{algorithm}
  
  \vspace{-2mm}
 \subsection{Optimize The Transmit Beamforming vector $\boldsymbol{\rm {w}}_{2,k}$}
 \label{subsec:a2}
 For the other variables given, the subproblem to optimize $\boldsymbol{\rm {w}}_{2,k}$ is redescribed as follows:
  \begin{subequations}\label{Problem:yhwt_new2}
 	\begin{alignat}{2}
 	\textbf{\textit{P}3-A:}
 	\mathop{\max}\limits_{\boldsymbol{\rm {w}}_{2,k}}&\ \frac{1}{r}\sum_{i=1}^{r}\sum_{k=1}^{K}R_{2,k}(\boldsymbol{\rm {w}}_{2,k};\vartheta_i) \notag\\
 	{\rm{s.t.}}:&1).\ \sum_{k=1}^{K}\Vert\boldsymbol{\rm {w}}_{2,k}\Vert^2\le P_{\max};\\
 	&2).\ E_{R} \ge \left(1-\tau\right)P_{R}(\boldsymbol{\rm {w}}_{2,k},\boldsymbol{\rm {\Theta}}).
 	\end{alignat}
 \end{subequations}
 Similar to the subproblem to optimize $\boldsymbol{\rm {w}}_{1,k}$, we apply the Quadratic Transform and Lagrangian Dual Transform to convert the objective function of $\textbf{\textit{P}3-A}$ by invoking  auxiliary variables $\boldsymbol{\rm {\omega}}_2 =\left[\omega_{2,1},\cdots,\omega_{2,K}\right]^T$ and $\boldsymbol{\rm {\nu}}_2=\left[\nu_{2,1},\cdots,\nu_{2,K}\right]^T$ and omitting irrelevant items as follows:
 
 \begin{small}\vspace{-5mm}
 	\begin{align}
 	&\hspace{-2mm}f_{OF}^{I\hspace{-0.5mm}I}\left(\boldsymbol{\rm {w}}_{2,k},\boldsymbol{\rm {\omega}}_2,\boldsymbol{\rm {\nu}}_2\right)\nonumber\\
 	=&\frac{1}{r}\hspace{-1mm}\sum_{i=1}^{r}\hspace{-1mm}\sum_{k=1}^{K}\hspace{-1mm}\left[\mbox{ln}(1\hspace{-1mm}+\hspace{-1mm}\omega_{2,k})\hspace{-1mm}+\hspace{-1mm}2\sqrt{1\hspace{-1mm}+\hspace{-1mm}\omega_{2,k}}\mbox{Re}\hspace{-1mm}\left\{\hspace{-1mm}\nu_{2,k}^*\boldsymbol{\rm {h}}_{k}^H\boldsymbol{\rm {w}}_{2,k}\hspace{-1mm}\right\}-\hspace{-1mm}\omega_{2,k}\hspace{-1mm}\right.\nonumber\\
 	&-\hspace{-1mm}\left.\left\vert\nu_{2,k}\right\vert^2\hspace{-1mm}\left(\hspace{-1mm}\textstyle\sum\limits_{j=1}^{K}\left\vert\boldsymbol{\rm {h}}_{k}^H\boldsymbol{\rm {w}}_{2,j}\right\vert^2\hspace{-2mm}+\hspace{-1mm}\sigma^2_{R}\left\Vert\boldsymbol{\rm {h}}_{RU,k}^H\boldsymbol{\rm {\Theta}}\right\Vert^2\hspace{-2mm}+\hspace{-1mm}Z_{2,k}(\vartheta_i)+\hspace{-1mm}\sigma^2_{2}\hspace{-1mm}\right)\hspace{-1mm}\right]\hspace{-1mm},
 	\end{align}
 \end{small}where $Z_{2,k}(\vartheta_i)$ represents $Z_{2,k}$ at $i$th channel realization. Thus, we have
  \begin{subequations}\label{Problem:yhwt2-1}
  	\begin{alignat}{2}
  		\textbf{\textit{P}3-B:}
  		&\mathop{\max}\limits_{\boldsymbol{\rm {w}}_{2,k},\boldsymbol{\rm {\omega}}_2,\boldsymbol{\rm {\nu}}_2} f_{OF}^{I\hspace{-1mm}I}\left(\boldsymbol{\rm {w}}_{2,k},\boldsymbol{\rm {\omega}}_2,\boldsymbol{\rm {\nu}}_2\right)
  		~{\rm{s.t.}}: (\ref{Problem:yhwt_new2}{\text{a}})~\text{and}~(\ref{Problem:yhwt_new2}{b})\nonumber.
  	\end{alignat}
  \end{subequations}

  Similarly, for all $k\in \mathcal{K}$, by letting $\partial f_{OF}^{I\hspace{-1mm}I}/\partial \omega_{2,k}$ and $\partial f_{OF}^{I\hspace{-1mm}I}/\partial\nu_{2,k}$ be zero, the optimal $\omega_{2,k}^{\star}$ and $\nu_{2,k}^{\star}$ are gotten, as follows:
  	\begin{align}\label{omage_2k}
  	\omega_{2,k}^{\star}\hspace{-1mm}=&\frac{\left\lvert\boldsymbol{\rm h}_{k}^H\boldsymbol{\rm {w}}_{2,k}\right\rvert^2}{\hspace{-3mm}\sum\limits_{j=1,j\neq k}^{K}\hspace{-1mm}\left\lvert\boldsymbol{\rm h}_{k}^H\boldsymbol{\rm {w}}_{2,j}\right\rvert^2\hspace{-3mm}+\hspace{-1mm}\sigma^2_{R}\hspace{-1mm}\left\Vert\boldsymbol{\rm {h}}_{RU,k}^H\boldsymbol{\rm {\Theta}}\right\Vert^2\hspace{-3mm}+\hspace{-1mm}\frac{1}{r}\hspace{-1mm}\sum\limits_{i=1}^{r}\hspace{-1.5mm}Z_{2,k}(\vartheta_i)\hspace{-1mm}+\hspace{-1mm}\sigma^2_{2}}\hspace{-1mm},\\
  	\nu_{2,k}^{\star}=&\frac{\sqrt{1-\omega_{2,k}}\boldsymbol{\rm h}_{k}^H\boldsymbol{\rm {w}}_{2,k}}{\sum\limits_{j=1}^{K}\hspace{-1mm}\left\vert\boldsymbol{\rm h}_{k}^H\boldsymbol{\rm {w}}_{2,j}\right\vert^2\hspace{-2mm}+\hspace{-1mm}\sigma^2_{R}\hspace{-1mm}\left\Vert\boldsymbol{\rm {h}}_{RU,k}^H\boldsymbol{\rm {\Theta}}\right\Vert^2\hspace{-3mm}+\hspace{-1mm}\frac{1}{r}\hspace{-1mm}\sum\limits_{i=1}^{r}\hspace{-1mm}Z_{2,k}(\vartheta_i)\hspace{-1mm}+\hspace{-1mm}\sigma^2_{2}}.\label{nu_2k}
  	\end{align}

  For brevity, we define that $\boldsymbol{\rm w}_{2}=[\boldsymbol{\rm w}_{2,1}^T,\cdots,\boldsymbol{\rm w}_{2,K}^T]^T$. Then, by omitting constant terms independent of $\boldsymbol{\rm {w}}_{2}$, the subproblem is further rewritten as follows:
  \begin{subequations}\label{Problem:yhwt2-2}
  	\begin{alignat}{2}
  		\textbf{\textit{P}3-C:}
  		&\mathop{\max}\limits_{\boldsymbol{\rm {w}}_2} \ \mbox{Re}\left\{\boldsymbol{\rm {y}}^H\boldsymbol{\rm {w}}_2\right\}-\boldsymbol{\rm {w}}_2^H\boldsymbol{\rm {Y}}\boldsymbol{\rm {w}}_2 \notag \\
  		{\rm{s.t.}}:&1).\ \boldsymbol{\rm {w}}_2^H\boldsymbol{\rm {w}}_2\le P_{\max};\\
  		&2).\ \boldsymbol{\rm {w}}_2^H\boldsymbol{\rm {S}}\boldsymbol{\rm {w}}_2\le P_E;
  	\end{alignat}
  \end{subequations}
  where we define 
  	\begin{align}
  	&\boldsymbol{\rm {y}}\triangleq[\boldsymbol{\rm {y}}_1^T,\cdots,\boldsymbol{\rm {y}}_K^T]^T,~\boldsymbol{\rm {y}}_k=2\sqrt{1+\omega_{2,k}}\nu_{2,k}\boldsymbol{\rm {h}}_k,\notag\\
  	&\boldsymbol{\rm {Y}}\triangleq\boldsymbol{\rm {I}}_k\otimes\left(\sum_{k=1}^{K}\vert \nu_{2,k}\vert^2\boldsymbol{\rm {h}}_k\boldsymbol{\rm {h}}_k^H\right),\notag\\
  	&\boldsymbol{\rm {S}}\triangleq\boldsymbol{\rm {I}}_k\otimes\left(\boldsymbol{\rm {G}}_{BR}^H\boldsymbol{\rm {\Theta}}\boldsymbol{\rm {\Theta}}\boldsymbol{\rm {G}}_{BR}\right),\notag\\
  	&P_E \triangleq\left(\left(1-\tau\right)\xi\right)^{-1}\left(E_R\hspace{-1mm}-\left(1-\tau\right)M(P_{dc}+P_{sc})\right.\nonumber\\
  	&\qquad\qquad\qquad\qquad\qquad\left.-\left(1-\tau\right)\xi\sigma^2_R\Vert\boldsymbol {\rm \Theta}\boldsymbol {\rm I}_M\Vert^2\right).
  	\end{align}
 $\textbf{\textit{P}3-C}$ is a convex problem that conforms to the criterion of disciplined convex programming (DCP), so we can apply convex solvers (CVX) to obtain the optimal solutions.

 \subsection{Optimize The Reflecting Beamforming  $\boldsymbol{\rm {\Theta}}$}
 \label{subsec:b}
 When the other variables are given, we proceed to study the optimization scheme for the reflecting beamforming  $\boldsymbol{\rm {\Theta}}$. The subproblem for $\boldsymbol{\rm {\Theta}}$ is recast as follows:
 \begin{subequations}\label{Problem:yhwt3-1}
 	\begin{alignat}{2}
 		\textbf{\textit{P}4-A:}
 		\mathop{\max}\limits_{\boldsymbol{\rm {\Theta}}} f_{OF}^{I\hspace{-1mm}I}\left(\boldsymbol{\rm {\Theta}}\right)
 		~{\rm{s.t.}}:~(\ref{Problem:yhwt1}{\text{b}})~\text{and}~ (\ref{Problem:yhwt1}{\text{c}})\nonumber.
 	\end{alignat}
 \end{subequations} 
 For ease of notation, we introduce the new definitions $e_{kj}\triangleq\boldsymbol{\rm {h}}_{BU,k}^H\boldsymbol{\rm {w}}_{2,j}$ and $\boldsymbol{\rm {\mu}}_j\triangleq\boldsymbol{\rm {G}}_{BR}\boldsymbol{\rm {w}}_{2,j}$, $\forall j\in\mathcal{K}$. Then, $\boldsymbol{\rm {h}}_{k}^H\boldsymbol{\rm {w}}_{2,j}$ can be redescribed as
 \begin{align}\label{gongshi1}
    \boldsymbol{\rm {h}}_{k}^H\boldsymbol{\rm {w}}_{2,j}=e_{kj}+\boldsymbol{\rm {h}}_{RU,k}^H\mbox{diag}\left(\boldsymbol{\rm {\mu}}_j\right)\boldsymbol{\rm {\theta}}, \forall k,j\in\mathcal{K}.
 \end{align}
 Similarly, $\boldsymbol{\rm {h}}_{J,qk}^H\left(\vartheta_i\right)\boldsymbol{\rm {z}}_{J,qk}$ can be reconstructed as
 \begin{align}\label{gongshi2}
 	\boldsymbol{\rm {h}}_{J,qk}^H\left(\vartheta_i\right)\boldsymbol{\rm {z}}_{J,qk}\hspace{-1mm}=\hspace{-1mm}d_{qk}\left(\vartheta_i\right)\hspace{-1mm}+\hspace{-1mm}\boldsymbol{\rm {h}}_{RU,k}^H\mbox{diag}\left(\boldsymbol{\rm {t}}_{qk}\left(\vartheta_i\right)\right)\boldsymbol{\rm {\theta}},
 \end{align}
 where the above newly introduced definitions are given as $d_{qk}\left(\vartheta_i\right)\triangleq\boldsymbol{\rm {h}}_{JU,qk}^H\left(\vartheta_i\right)\boldsymbol{\rm {z}}_{J,qk}$ and $\boldsymbol{\rm {t}}_{qk}\left(\vartheta_i\right)\triangleq\boldsymbol{\rm {G}}_{JR,q}\left(\vartheta_i\right)\boldsymbol{\rm {z}}_{J,qk}$. 
 
 By substituting Eq.~(\ref{gongshi1}) and Eq.~(\ref{gongshi2}) and dropping the irrelevant
 terms with respect to $\boldsymbol{\rm {\theta}}$, the subproblem $\textbf{\textit{P}4-A}$ is reconstructed as follows:
 \begin{subequations}\label{Problem:yhwt41}
	\begin{alignat}{2}
		\textbf{\textit{P}4-B:}
		\mathop{\max}\limits_{\boldsymbol{\rm {\theta}}}&~\frac{1}{r}\sum_{i=1}^{r}\left(\mbox{Re}\left\{\boldsymbol{\rm {\theta}}^H\boldsymbol{\rm {\Lambda}}\left(\vartheta_i\right)\right\}-\boldsymbol{\rm {\theta}}^H\boldsymbol{\rm {\Gamma}}\left(\vartheta_i\right)\boldsymbol{\rm {\theta}}\right) \notag\\
		{\rm{s.t.}}:&1).\  \boldsymbol{\rm {\theta}}^H\boldsymbol{\rm {V}}\boldsymbol{\rm {\theta}} \le \widetilde{P}_E;\\
		&2).\ A_{m} \le A_{\max},\forall m \in \mathcal{M},
	\end{alignat}
\end{subequations}
where the parameters are defined as follows:
	\begin{align}
		&\boldsymbol{\rm {\Gamma}}\left(\vartheta_i\right)\nonumber\\
		&\triangleq\sum_{k=1}^{K}\left\vert\nu_{2,k}\right\vert^2\left(\sum_{q=1}^{Q}\mbox{diag}\left(\boldsymbol{\rm {t}}^*_{qk}\left(\vartheta_i\right)\right)\boldsymbol{\rm {h}}_{RU,k}\boldsymbol{\rm {h}}_{RU,k}^H\mbox{diag}\left(\boldsymbol{\rm {t}}_{qk}\left(\vartheta_i\right)\right)\right.\nonumber\\
		&+\hspace{-2mm}\left.\sum_{j=1}^{K}\hspace{-1mm}\mbox{diag}\hspace{-1mm}\left(\boldsymbol{\rm {\mu}}^*_{j}\right)\hspace{-1mm}\boldsymbol{\rm {h}}_{RU,k}\boldsymbol{\rm {h}}_{RU,k}^H\mbox{diag}\hspace{-1mm}\left(\boldsymbol{\rm {\mu}}_{j}\right)\hspace{-1mm}+\hspace{-1mm}\sigma^2_R\mbox{diag}\left(\boldsymbol{\rm {h}}_{RU,k}\hspace{-1mm}\odot\hspace{-1mm}\boldsymbol{\rm {h}}_{RU,k}^*\right)\hspace{-1mm}\right)\hspace{-1mm},\nonumber\\
		&\boldsymbol{\rm {\Lambda}}\left(\vartheta_i\right)
		\triangleq\sum_{k=1}^{K}\mbox{diag}\left(\boldsymbol{\rm {h}}_{RU,k}^*\right)\left(2\sqrt{1\hspace{-1mm}+\hspace{-1mm}\omega_{2,k}}\nu_{2,k}^*\boldsymbol{\rm {\mu}}_k\right.\nonumber\\
		&\qquad\qquad-\left.\left\vert\nu_{2,k}\right\vert^2\sum_{j=1}^{K}e_{kj}^*\boldsymbol{\rm {\mu}}_j-\left\vert\nu_{2,k}\right\vert^2\sum_{q=1}^{Q}d_{qk}^*\left(\vartheta_i\right)\boldsymbol{\rm {t}}_{qk}\left(\vartheta_i\right)\right),\nonumber\\
		&\boldsymbol{\rm {V}}\triangleq\sum_{k=1}^{K}\mbox{diag}\left(\boldsymbol{\rm {\mu}}_{k}\hspace{-1mm}\odot\hspace{-1mm}\boldsymbol{\rm {\mu}}_{k}^*\right)+\sigma_R^2\boldsymbol{\rm {I}}_{M},\nonumber\\
		&\widetilde{P}_E\triangleq\left(\left(1-\tau\right)\xi\right)^{-1}\left(E_R\hspace{-1mm}-\left(1-\tau\right)M(P_{dc}+P_{sc})\right).
	\end{align}
Hence, the subproblem $\textbf{\textit{P}4-B}$ is convex, so we can get the optimal solution for $\textbf{\textit{P}4-B}$ by adopting the standard convex solver.

\begin{algorithm}[t]
	\caption{ SSCA-based AO Algorithm for Problem \textbf{\textit{P}1}.}
	\label{alg:Framwork}
	\begin{algorithmic}[1]
		\Require
		Initialize $\boldsymbol{\rm {w}}_{1,k}^{(0)}$, $\boldsymbol{\rm {w}}_{2,k}^{(0)}$, $\boldsymbol{\rm {\Theta}}^{(0)}$, $\tau^{(0)}$,the number of iterations $r=0$, the accuracy $\varsigma=10^{-3}$, and the maximum iterations $r_{\max}$.
		\Ensure
		\State Obtain new channel realizations $\boldsymbol{\rm h}_{JU,qk}(\vartheta_r)$, $\boldsymbol{\rm h}_{IU,bk}(\vartheta_r)$, $\boldsymbol{\rm G}_{JR,q}(\vartheta_r)$.
		\State Compute the objective function value of $\textbf{\textit{P}1}$, denoted by $V_{OF}\left(r\right)$.
		\State Update $\tau^{(r+1)}$ by (\ref{tau}).
		\State Update $\boldsymbol{\rm {\omega}}_{1}^{(r+1)}$ and $\boldsymbol{\rm {\nu}}_2^{(r+1)}$ by (\ref{omage_k}) and (\ref{nu_k}), respectively.
		\State Tackle the subproblem $\textbf{\textit{P}2-C}$ and get the optimal solution $\boldsymbol{\rm {w}}_{1,k}^{(r+1)}$ by utilizing Algorithm~\ref{alg:SCA_erfen}.
		\State Calculate $\boldsymbol{\rm {\omega}}_{2}^{(r+1)}$ and $\boldsymbol{\rm {\nu}}_2^{(i+1)}$ by (\ref{omage_2k}) and (\ref{nu_2k}), respectively.
		\State Tackle the subproblem $\textbf{\textit{P}3-C}$ and get the optimal solution $\boldsymbol{\rm {w}}_{2,k}^{(r+1)}$.
		\State Calculate the optimal solution $\boldsymbol{\rm {\Theta}}^{(r+1)}$ by solving the subproblem $\textbf{\textit{P}4-A}$.
		\State Update $r=r+1$.
		\renewcommand{\algorithmicrequire}{\textbf{until}}
		\Require $\left\vert V_{OF}\left(r+1\right)-V_{OF}\left(r\right)\right\vert/V_{OF}\left(r+1\right)<\varsigma$ or $r = r_{\max}$.
	\end{algorithmic}
\end{algorithm}

 The SSCA-based AO algorithm for solving $\textbf{\textit{P}1}$ is presented in Algorithm \ref{alg:Framwork}. After derivation, all subproblems in our proposed algorithm are written as convex problems, which can be tackled by SCA algorithm and the interior-point-method (IPM) based CVX toolbox. Consequently, the complexity of Algorithm \ref{alg:Framwork} is expressed as $\mathcal{O}\left(r_{\max}\left(M^{3.5}+(NK)^{3.5}+i_{\max}KN^3\mbox{log}_2\frac{\lambda_{U}-\lambda_{L}}{2}\right)\right)$. According to the algorithm mentioned above, the algorithm complexity grows with the increase of REs at the active RIS and also leads to the rise in hardware complexity \cite{Aided-Wireless}. Therefore, it is very crucial to configure the number of REs, when planning anti-jamming scheme with the aid of active RIS.

 \vspace{-2mm}
\section{Numerical Simulations and Analysis}
\label{sec:simulation}

\begin{figure}[t!]
 	\vspace{-12pt}
 	\centering
 	\includegraphics[scale=0.065]{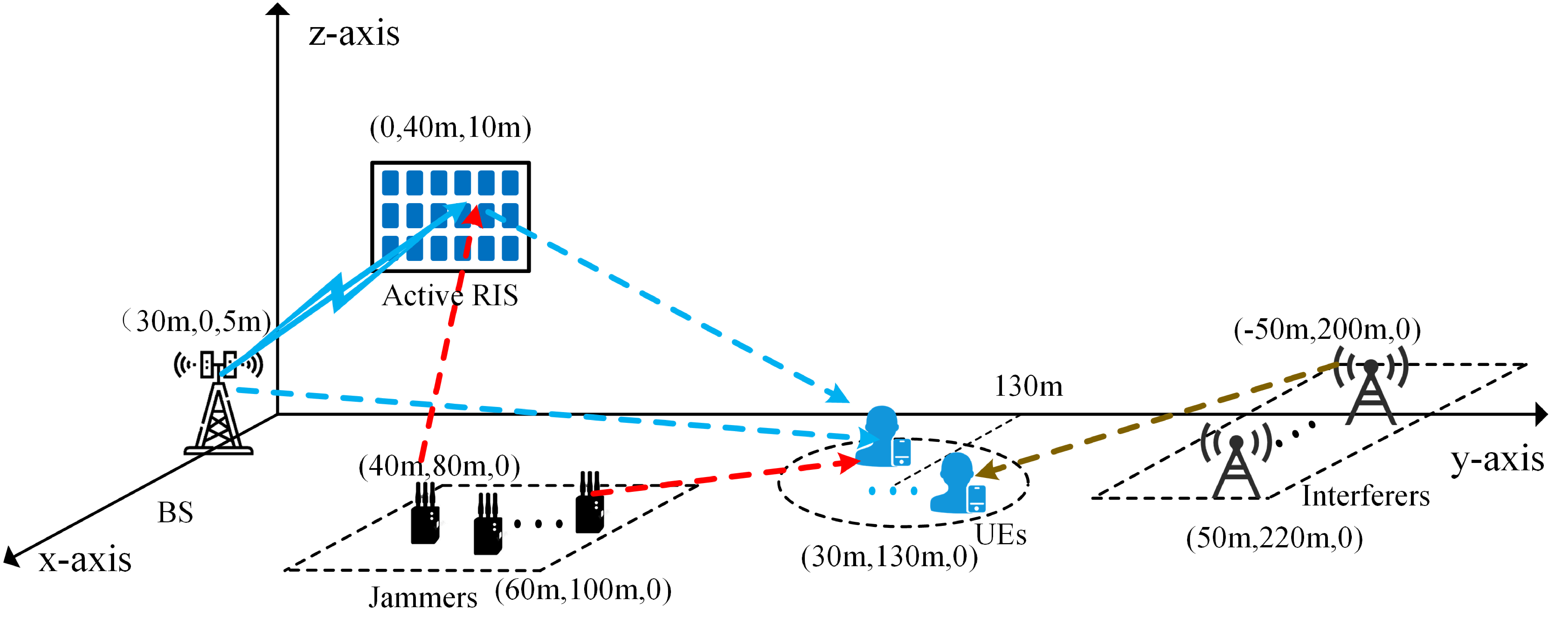}
 	\caption{Plane diagram of simulated active RIS-assisted anti-jamming system.}
 	\vspace{-15pt}
 	\label{fig:simulation1}
 \end{figure}

 During this section, we proceed with simulations experiments to assess the performance of our developed BS harvesting scheme for the self-sustainable active RIS-assisted anti-jamming system. Unless otherwise stated, the setting of parameters runs throughout our simulations. In the system, the BS with $N = 8$ antennas provides communication services to $K = 4$ UEs in the presence of $Q = 3$ malicious jammers and $B = 4$ interferers, where the
 antennas number of each jammer and each interferer are set as $N_{Jam} = 8$ and $N=8$, respectively. In addition, the jamming and the interference power are given as $P_{J}\hspace{-1mm} =\hspace{-1mm}P_{J,q}\hspace{-1mm} =\hspace{-1mm}\sum_{k=1}^{K}\lVert\boldsymbol{\rm {z}}_{J,qk}\rVert^2 \hspace{-1mm}=\hspace{-1mm}10 $ dBm and $P_{I}\hspace{-1mm} =\hspace{-1mm}P_{I,b}\hspace{-1mm} =\hspace{-1mm}\sum_{k=1}^{K}\lVert\boldsymbol{\rm {z}}_{I,bk}\rVert^2 \hspace{-1mm}=\hspace{-1mm}10 $ dBm. The power of noise introduced by UEs and active RIS is all set as -105 dBm. Besides, we set the control circuit power and the DC biasing power for each RE as $P_{sc}=10$ $\mu W$, $P_{dc}=10$ $\mu W$, respectively. According to \cite{Aided-Wireless}, we establish a coordinate system as in Fig.~\ref{fig:simulation1}, where (30m, 0, 5m) and (0, 40m, 10m) are the coordinates of the BS and active RIS, respectively. The movement range of the legitimate UEs is covered by an area centered on (30m, 150m, 0) with a radius of 20m. The range of jammers appearance is restricted by the 20 by 20 meters area, determined by (40m, 80m, 0), and (60m, 100m, 0). Regarding interferers that contribute to co-channel interference, they are set within a rectangular region constrained by the boundaries defined by (-50m, 200m, 0) and (50m, 220m, 0). Based on the above settings, we can get the Euclidean distance among BS, active RIS, UEs, jammers and interferers by simple algebraic operation. Values corresponding to three dimensional (3D) topological structure are adopted as values of RWP parameters ($B_n=[735/72, -1190/72, 455/72]$, $\Upsilon_n=[1,3,5]$, and $N_T=3$) \cite{Cao1}. The Nakagami-$m$ parameter is $m_N=1$. 
 
 The path loss of the wireless channel in dB is stated as $PL =  \zeta_0 +10\alpha_{pl} \mbox{log}_{10}(d/d_0)$, where $\alpha_{pl}$ and $\zeta_0$ represent the path loss exponent and the path loss for the reference distance $d_0=1$ m. According to \cite{Joint-Active}, we use $\alpha_{BU} = 2.75$, $\alpha_{BR} = 2.2$, $\alpha_{RU} = 2.2$, $\alpha_{JU} = 2.5$, $\alpha_{JR} = 2.5$ and $\alpha_{IU} = 2.7$ to indicate the path loss exponents corresponding to the channel $\boldsymbol{\rm h}_{BU,k}$, $\boldsymbol{\rm G}_{BR}$, $\boldsymbol {\rm h}_{RU,k}$, $\boldsymbol {\rm h}_{JU,qk}$, $\boldsymbol {\rm G}_{JR,q}$, and $\boldsymbol {\rm h}_{IU,bk}$, respectively. In response to imperfect CSI, we require estimates of the channels of jammers and interferers, i.e., $\boldsymbol{\rm {h}}_{JU,qk}$ and $\boldsymbol{\rm {G}}_{JR,q}$, and $\boldsymbol{\rm {h}}_{IU,bk}$. Denote one of the above channels as $h_x$, and the corresponding channel estimate as $\widehat{h}_x$. We further consider that the estimation error $h_x-\widehat{h}_x$ follows the complex Gaussian distribution with zero mean. Besides, all channels have the mean-square error (MSE), as as $e_{MSE}=\mathbb{E}\left[\vert h_x-\widehat{h}_x\vert^2\right]/\mathbb{E}\left[\vert \widehat{h}_x\vert^2\right]$.All simulation curves are obtained by averaging 500 random tests.
 


For the purposes of this comparative analysis, we evaluate our developed scheme against the  below schemes: \textbf{The passive RIS scheme}: the conventional AO algorithm is employed to optimize relevant variables, thereby strengthening the anti-jamming performance of traditional passive RIS aided system; \textbf{The scheme without RIS}: the traditional AO algorithm is adopted to optimize relevant variables, thereby strengthening the anti-jamming performance of system without RIS. 

 \begin{figure}[t]
	\vspace{-20pt}
	\centering
	\includegraphics[scale=0.5]{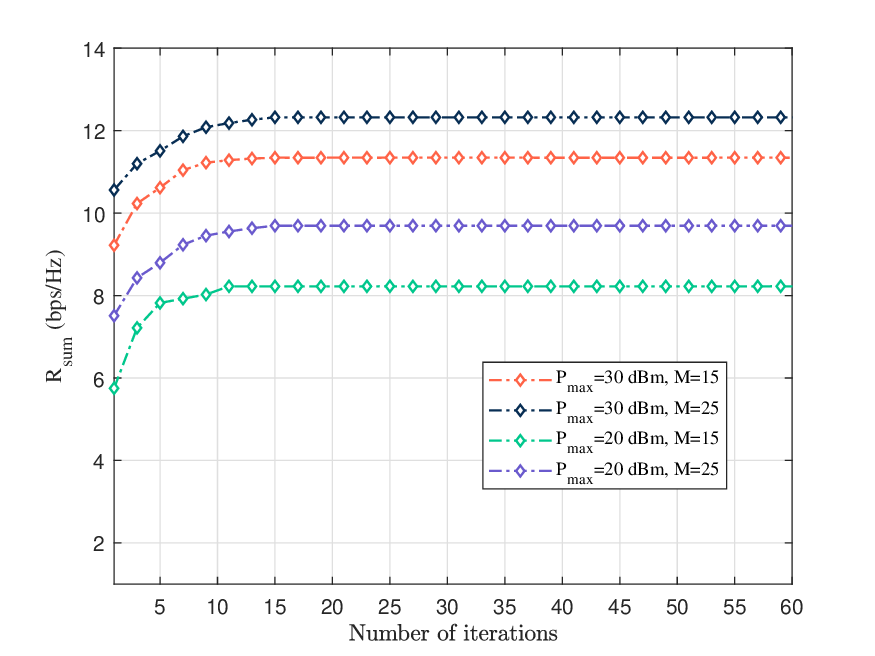}
	\vspace{-10pt}
	\caption{Achievable rate $R_{sum}$ versus the number of iterations.}
	\vspace{-13pt}
	\label{fig:ite_R}
\end{figure}

  The convergence of SSCA-based AO algorithm of the developed BS harvesting scheme for anti-jamming system is presented in Fig.~\ref{fig:ite_R}. Assume that the maximum amplification gain is $A_{\max}^2=40$ dB \cite{zhengming}. The achievable rate $R_{sum}$ of the system first significantly increases, and then keeps unchanged within $15$ iterations, when the number of iterations increases. 
 
 \begin{figure}[t]
	\centering
	\includegraphics[scale=0.50]{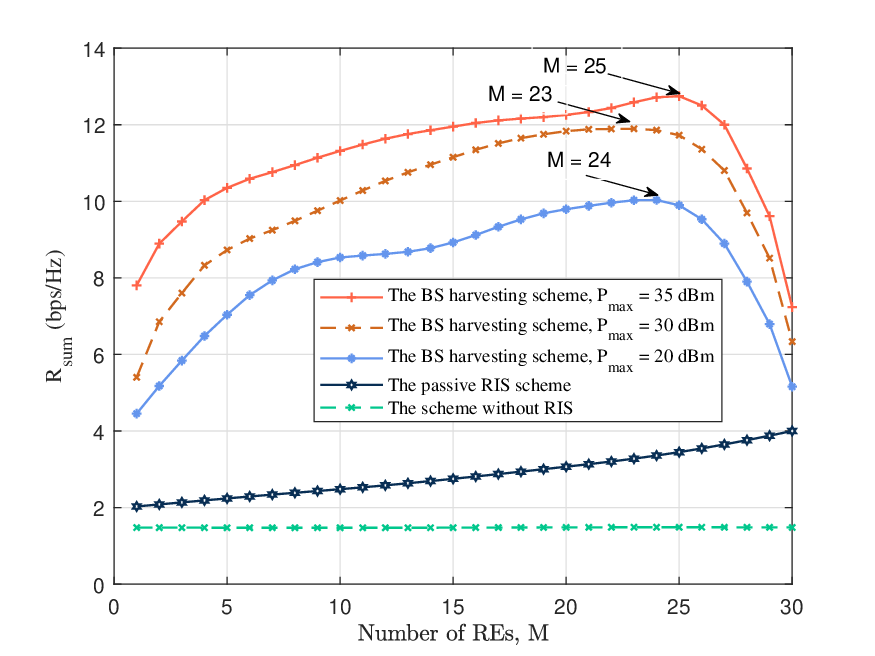}
	\vspace{-10pt}
	\caption{Achievable rate $R_{sum}$ versus the number of REs $M$.}
	\vspace{-15pt}
	\label{fig:M_R}
\end{figure}

 Figure~\ref{fig:M_R} compares the achievable rate $R_{sum}$ with the number of REs at RISs for different schemes in anti-jamming systems. We can see that achievable rates for passive RIS-aided system and the system without RIS are substantially lower than that for active RIS-assisted system since the active RIS magnifies amplitude of the signal to weaken the influence of double-fading attenuation. As $M$ increases, achievable rate curve of the active RIS-assisted system initially rises and falls, which leads to an optimal point when $M\hspace{-1mm} = \hspace{-1mm}25$. This result is caused by the trade-off between the number of REs at active RIS and the transmit power budget in our designed BS harvesting scheme. In a given transmit power budget, the anti-jamming performance can benefit more from the initial increase of $M$, because the amplification power of a small number of REs to reflect signals is not limited by the transmit power budget $P_{\max}$. Then, when the number of REs grows, more transmit power is used for power supply in our structure. Therefore, due to the limitation of transmit power budget, more REs reduce the power used for amplification after exceeding the optimal point. In addition, BS harvesting scheme can achieve sufficiently high performance gain when deploying a small number of REs, which provides the feasibility for deploying the active RIS in a limited area. 

 \begin{figure}[t]
 	\vspace{-20pt}
 	\centering
 	\includegraphics[scale=0.5]{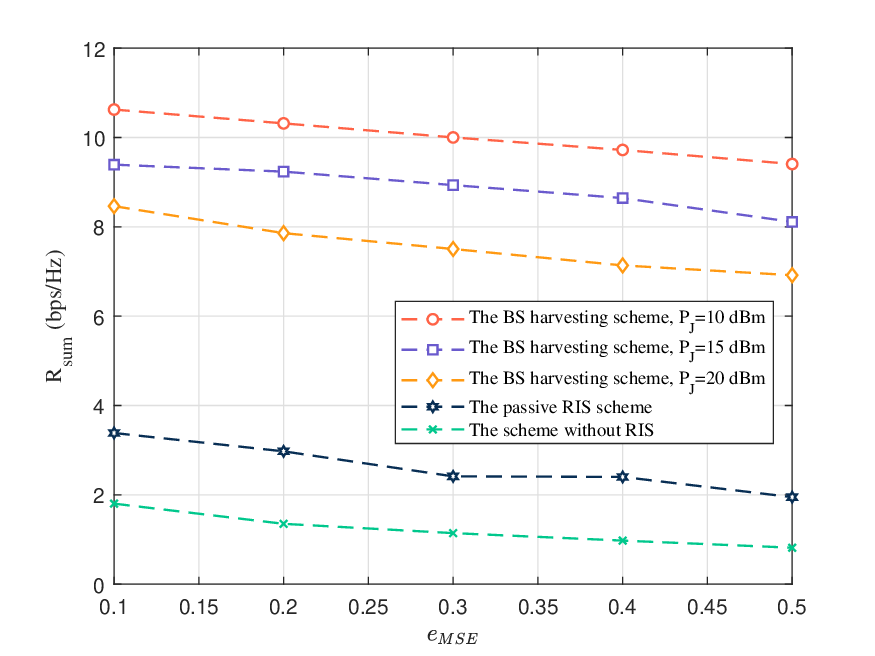}
 	\vspace{-10pt}
 	\caption{Achievable rate $R_{sum}$ versus the normalized MSE $e_{MSE}$ of estimated channels.}
 	\vspace{-13pt}
 	\label{fig:e_R}
 \end{figure}

  Figure~\ref{fig:e_R} compares the achievable rate $R_{sum}$ with the normalized MSE $e_{MSE}$ of estimated channels under different schemes. We can see that as $e_{MSE}$ increases, the lower the achievable rate obtained by all the schemes. The reason for this is that as $e_{MSE}$ grows, the estimated channels are more uncertain, leading to a lower accuracy of the optimized beamforming at the BS and RIS, which makes the obtained achievable rate lower. Furthermore, as expected, we can see that the higher the jamming power $P_J$, the lower the achievable rate obtained.

\begin{figure}[t]
 	\centering
 	\includegraphics[scale=0.5]{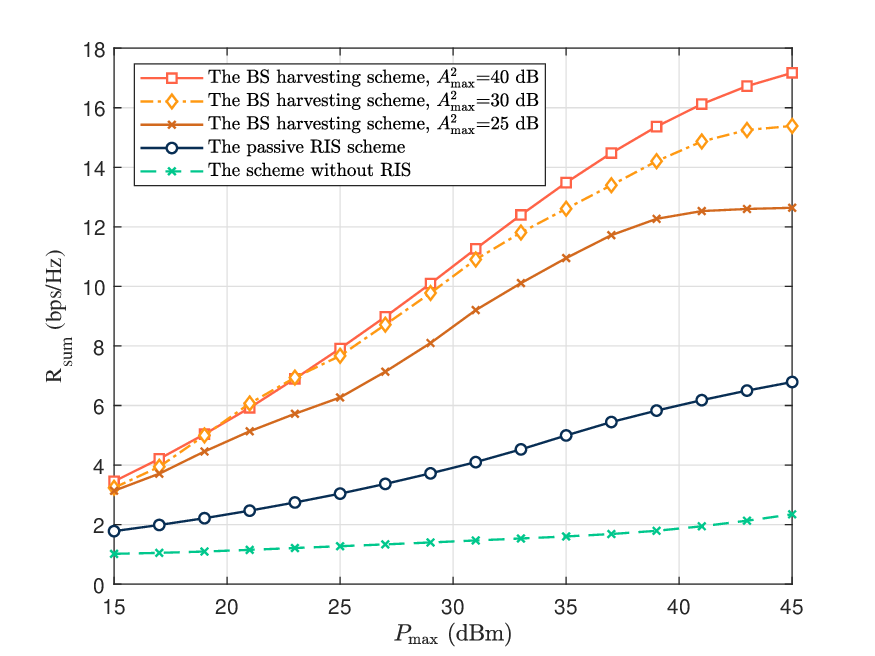}
 	\vspace{-10pt}
 	\caption{Achievable rate $R_{sum}$ versus the maximum transmit power budget $P_{\max}$, when $M=25$.}
 	\vspace{-10pt}
 	\label{fig:Pmax_R}
 \end{figure}

 Figure~\ref{fig:Pmax_R} compares the achievable rate $R_{sum}$ with maximum transmit power budget $P_{\max}$ using different schemes. With the continuous growth of the maximum transmit power $P_{\max}$, achievable rates $R_{sum}$ of all schemes monotonically increases. The BS harvesting scheme benefits more from the growth of $P_{\max}$ as compared with the other schemes. This is because that the influence of double-fading attenuation can be significantly weakened by active RIS, thus obtaining higher achievable rate. Our proposed self-sustainable active RIS uses the energy transmitted by the BS, which increases the deployment flexibility of the active RIS and reduces the hardware complexity. Fig.~\ref{fig:Pmax_R} also depicts $R_{sum}$ of the BS harvesting scheme under different amplification gain budgets $A_{\max}^2=20, 30$ and $40$ dB. The results demonstrate that $R_{sum}$ of the system with different $A_{\max}^2$ are almost the same when the maximum transmit power is low. The growth of $R_{sum}$ slows down, when the maximum transmit power grows. It indicates that the amplification power constraint is not always active. This constraint is invalid when $P_{\max}$ is small. With the increase of $P_{\max}$, the constraint becomes active, resulting in the reduction of the benefit of the BS harvesting scheme from the transmit power. Therefore, $R_{sum}$ of system does not continuously increase with the growth of $P_{\max}$. In addition, Fig.~\ref{fig:Pmax_R} shows that the anti-jamming performance of the system is better when the active RIS has a significant amplification gain budget of $A_{\max}^2$.  Therefore, it is essential to consider the trade-off between transmit power budget and amplification power budget when designing active RIS-assisted system.
 
\begin{figure}[t]
	\vspace{-20pt}
	\centering
	\includegraphics[scale=0.5]{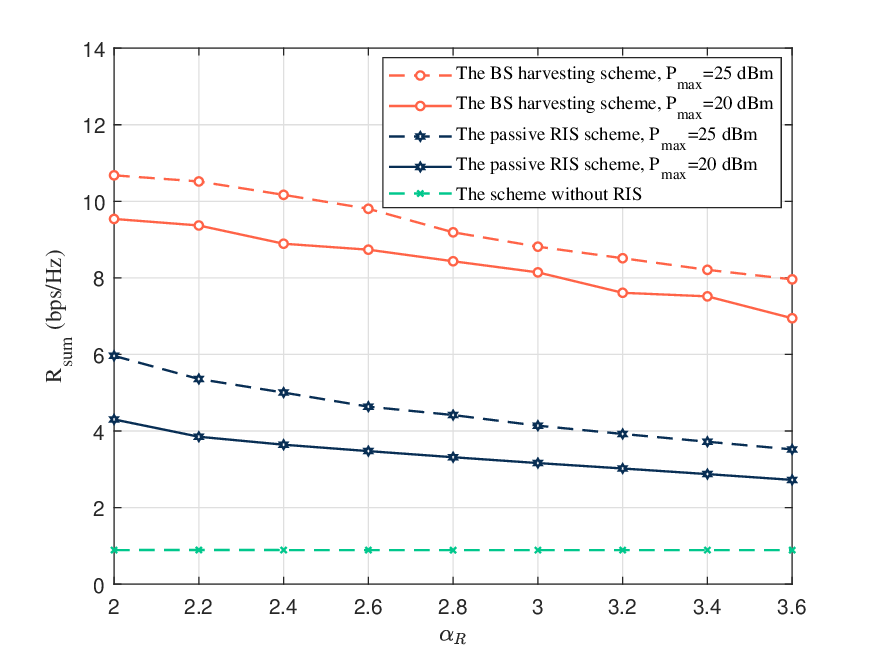}
	\vspace{-10pt}
	\caption{Achievable rate $R_{sum
		}$ versus the RIS-related path loss exponent $\alpha_{R}$.}
	\vspace{-13pt}
	\label{fig:alpha_R}
\end{figure}

 Figure~\ref{fig:alpha_R} compares the achievable rate $R_{sum}$ with the RIS-related path loss exponent $\alpha_{R}$, where $\alpha_{R} = \alpha_{BR} = \alpha_{RU}$. Our above results are based on the ideal channel. However, the fading of the channel constantly changes in practice, and the ideal situation is rare. Therefore, it is essential to analyze the influence of path loss exponent on anti-jamming performance. Clearly, $R_{sum}$ of the system using RIS (including active and passive RIS) decreases as $\alpha_{R}$ increases. The increase of path loss exponent means the deterioration of RIS-related channel, resulting in the system benefiting less from active RIS. To maintain higher performance, the active RIS is better set in the wireless environment with a lower $\alpha_{R}$.

\begin{figure}[t]
	\vspace{5pt}
	\centering
	\includegraphics[scale=0.5]{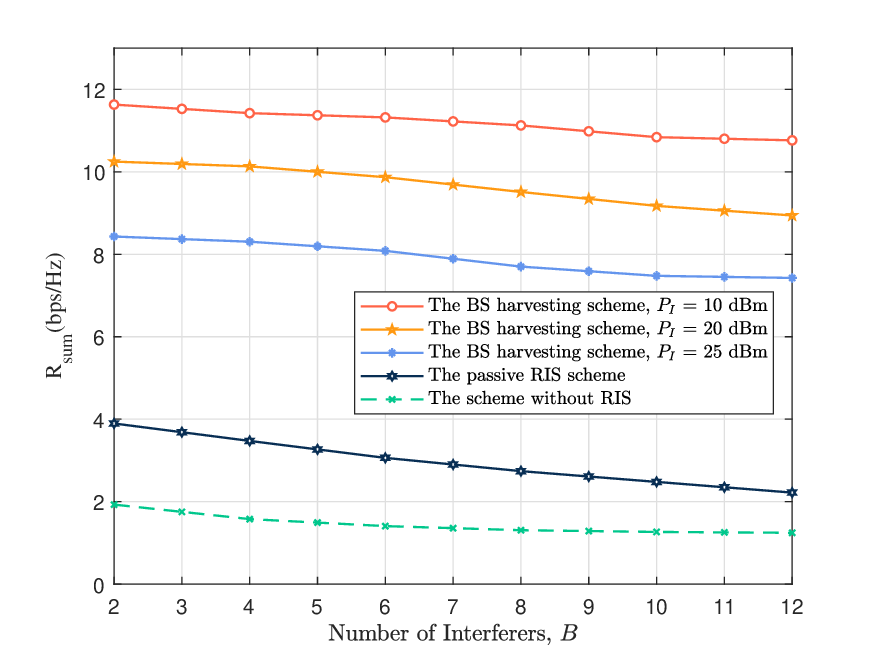}
	\vspace{-10pt}
	\caption{Achievable rate $R_{sum}$ versus the number of interferers $B$.}
	\vspace{-13pt}
	\label{fig:B_R}
\end{figure}
 
 
 Figure~\ref{fig:B_R} compares the achievable rate $R_{sum}$ with the number $B$ of interferers. As $B$ increases, the achievable rate of all schemes declines. The reason is that the co-channel interference becomes severe with the growth of $B$, leading to damage to legitimate communication. The achievable rate $R_{sum}$ of our developed BS harvesting scheme is significantly higher than that of other schemes. This is because our proposed scheme can effectively cancel co-channel interference. Therefore, we conclude that the active RIS can effectively mitigate the influence of legitimate communication in interference-limited scenarios. Moreover, we study the influence of interference power on $R_{sum}$ of the system. With the growth of $P_I$, $R_{sum}$ decreases inversely. As $P_I$ grows, the communication environment of the UEs becomes more hostile, thus inhibiting $R_{sum}$ of the communication system.



\section{Conclusions}\label{sec:Conclusion}
 In this paper, we studied the active RIS-assisted anti-jamming communication system, where the active RIS is utilized to resist malicious jamming from jammers and co-channel interference from interferers. We designed a novel self-sustainable structure in the system, where the active RIS is energized by harvesting energy from signals at the BS through TD-SWIPT scheme to reduce hardware complexity and enhance deployment flexibility. Based on the structure, we proposed the BS harvesting scheme that aims to maximize the achievable rate with joint transmit beamforming at the BS and reflecting beamforming at the active RIS. We also  tackled the nonconvex problem in the scheme by adopting the SSCA-based AO algorithm. In addition, the RWP based Nakagami-$m$ fading channel model was derived to simulate the related channels of UEs in the anti-jamming system. Simulation results show that under the same maximum transmit power, our developed BS harvesting scheme outperforms other schemes in the aspect of the anti-jamming performance when fewer REs are deployed. Also, we analyze the influence of the number of interferers, the normalized MSE of estimated channels, and path loss exponent on the active RIS-assisted system to validate the benefits of utilizing self-sustainable active RIS to enhance wireless communication performance in jamming assault scenarios.

\bibliographystyle{IEEEtran}
\bibliography{References}

\end{document}